\documentclass{article}

\usepackage[nonatbib,final]{neurips_2024}

\makeatletter
\renewcommand{\@notice}{}
\makeatother

\usepackage[utf8]{inputenc} %
\usepackage[T1]{fontenc}    %
\usepackage{hyperref}       %
\usepackage{url}            %
\usepackage{booktabs}       %
\usepackage{amsfonts}       %
\usepackage{nicefrac}       %
\usepackage{microtype}      %
\usepackage[table]{xcolor}         %
\usepackage{longtable}
\usepackage{graphicx}
\usepackage{float}
\usepackage{array}
\usepackage{soul}
\usepackage{titlesec} %
\usepackage{enumitem} %
\usepackage{caption}
\counterwithin{subsubsection}{subsection}
\renewcommand{\thesubsubsection}{\roman{subsubsection}}
\titleformat{\subsubsection}[hang]{\normalfont\bfseries}{\thesubsubsection.}{0.5em}{}[]
\titlespacing{\subsubsection}{0pt}{1.5ex plus .1ex minus .1ex}{1ex plus .1ex minus .1ex}

\definecolor{fillingblue}{HTML}{e6eef4}
\definecolor{blankcolor}{HTML}{4A4238}
\definecolor{blankcolor}{HTML}{700548}
\definecolor{blankcolor}{HTML}{2A0C4E}
\definecolor{blankcolor}{HTML}{45062E}

\newcommand{\blank}[1]{\sethlcolor{fillingblue}%
  {\color{blankcolor}\hl{ #1 }}}

\counterwithin{paragraph}{subsubsection}
\renewcommand{\theparagraph}{\thesubsubsection-\alph{paragraph}.}
\titleformat{\paragraph}[hang]{\normalfont\bfseries}{\theparagraph}{0.5em}{}[]
\titlespacing{\paragraph}{0pt}{1.5ex plus .1ex minus .1ex}{1ex plus .1ex minus .1ex}

\usepackage[backend=biber, style=apa, natbib]{biblatex}
\AtEveryBibitem{\clearfield{note}}
\AtEveryBibitem{%
  \ifentrytype{book}{%
    \clearfield{url}%
    \clearfield{urldate}%
  }{}%
}
\DeclareSourcemap{
  \maps[datatype=bibtex]{
    \map{
       \step[fieldsource=url,
             notmatch=\regexp{wiki},
             final=1]
       \step[fieldset=urldate, null]
    }
  } 
}
\usepackage{xurl}
\hypersetup{
    colorlinks=true,
    linkcolor=blue,
    citecolor=blue,
    filecolor=magenta,      
    urlcolor=cyan!60!black,
    }

\usepackage[most]{tcolorbox}
\usepackage{enumerate}
\newcommand{\examplebox}[2]{%
\definecolor{background}{HTML}{F4F4F6}

\begin{tcolorbox}[
    parbox=false,
    toptitle=8pt,
    colback=background,
    coltext=black,               
    colframe=black,         
    boxrule=0.8pt,               
    sharp corners,               
    colbacktitle=background, 
    coltitle=black,              
    fonttitle={\bfseries\centering}, 
    breakable,   
    title={#1},
    titlerule = -1pt,  
]
    #2
\end{tcolorbox}}

\setlist[itemize]{leftmargin=*}

\usepackage{cleveref}

\title{STREAM (ChemBio):\\ A Standard for Transparently Reporting \\Evaluations in AI Model Reports}

\let\makecite\cite
\newcommand{\uline}{}

\author{%
  Tegan McCaslin
  \\
  Independent 
  \And
  Jide Alaga \\
  METR \\
  \And
  Samira Nedungadi \\
  SecureBio \\
  \And
  Seth Donoughe\\
  SecureBio\\
  \AND
  Tom Reed\\
  GovAI\\
  \And
  Rishi Bommasani\\
  Stanford HAI\\
  \And 
  Chris Painter\\
  METR\\
  \And   
  Luca Righetti\thanks{Corresponding author: \texttt{{luca.righetti@governance.ai}}} \\
  GovAI
}

\begin{document}
  \newgeometry{
    textheight=9.2in,
    textwidth=5.5in,
    top=1in,
    headheight=12pt,
    headsep=25pt,
    footskip=30pt
  }\maketitle

\begin{abstract}
  Evaluations of dangerous AI capabilities are important for managing catastrophic risks. Public transparency into these evaluations—including what they test, how they are conducted, and how their results inform decisions—is crucial for building trust in AI development. We propose STREAM (A Standard for Transparently Reporting Evaluations in AI Model Reports), a standard to improve how model reports disclose evaluation results, initially focusing on chemical and biological (ChemBio) benchmarks. Developed in consultation with 23 experts across government, civil society, academia, and frontier AI companies, this standard is designed to (1) be a practical resource to help AI developers present evaluation results more clearly, and (2) help third parties identify whether model reports provide sufficient detail to assess the rigor of the ChemBio evaluations. We concretely demonstrate our proposed best practices with “gold standard” examples, and also provide a three-page reporting template to enable AI developers to implement our recommendations more easily.
\end{abstract}

\begin{figure}[H]
    \centering
    \includegraphics[width=0.65\linewidth]{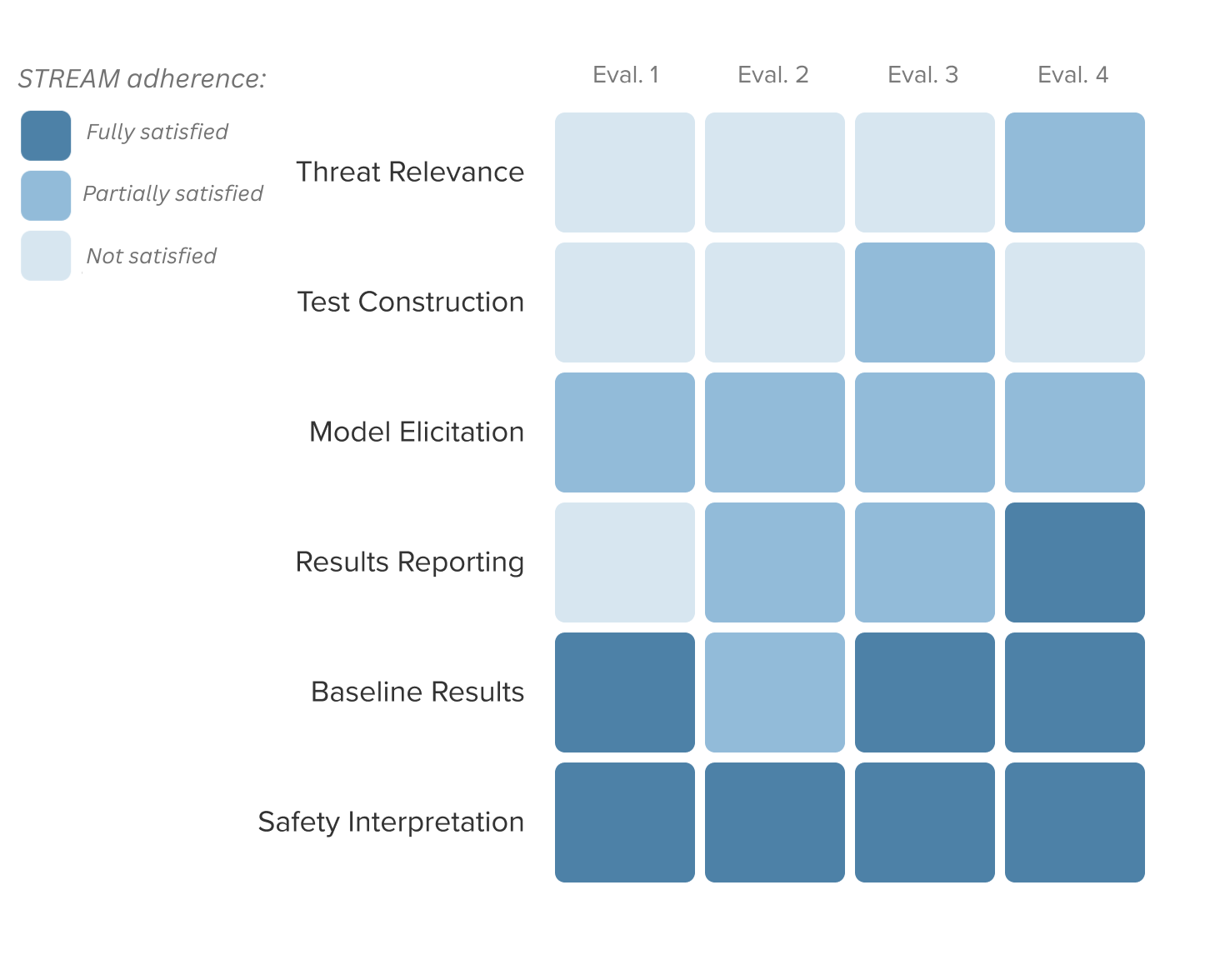}
    \caption{A stylized example of a model report graded using STREAM v1}
    \label{fig:STREAMreport}
\end{figure}

  \newgeometry{
    textheight=9in,
    textwidth=5.5in,
    top=1in,
    headheight=12pt,
    headsep=25pt,
    footskip=30pt
  }\section{Introduction\label{introduction}}

Powerful AI systems could provide great benefits to society, but may also bring large-scale risks \citep{ISRSAA2025,2024bc}, such as misuse of these systems by malicious actors \citep{mouton2024,brundage2018}. In response many leading AI companies have committed to regularly testing their systems for dangerous capabilities, including capabilities related to chemical and biological misuse \citep{metr2025a,forum2024a}. These tests are often referred to as ``dangerous capability evaluations'', and they are a key component of AI companies' Frontier AI Safety Policies (FSPs), voluntary commitments to the US White House \citep{thewhitehouse2023}, and the EU General-Purpose AI Code of Practice \citep{europeancommission2025}.

Despite their importance, there are currently no widely used standards for documenting dangerous capability evaluations clearly alongside model deployments \citep{weidinger2025, paskov2025c}. Several leading AI companies and AI Safety/Security Institutes regularly publish dangerous capability evaluation results in ``model reports'' (also called ``model cards'', ``system cards'' or ``safety cards'') and cite those results to support important claims about a models' level of risk \citep{mitchell2019}.\footnote{See, for example: \cite{zotero-3128}; \cite{google2025}; \cite{openai2025}; and \cite{aisecurityinstitute2024}.} But there is little consistency across such model reports on the evaluation details they provide.

In particular, many model reports lack sufficient information about how their evaluations were conducted, the output of the evaluations, and how the results informed judgments of a model's potentially dangerous capabilities \citep{reuel2024c,righetti2024,bowen2025a}. This limits how informative and credible any resulting claims can be to readers, and impedes third party attempts to replicate such results.\footnote{Note that even a highly detailed model report may not enable replication of all results, as they may involve private versions of models, scaffolding, or evaluations.}

We aim to facilitate better dangerous capability evaluation reporting by providing a clear and standardized reporting framework: a Standard for Transparently Reporting Evaluations in AI Model Reports (STREAM). This framework details the key information we view as necessary for AI developers to present results from dangerous capability evaluations more clearly,  allowing third parties to understand and interpret these results. Note that STREAM addresses the quality of \emph{reporting}, not the quality of the underlying evaluations or any resulting risk interpretations.

In this paper, we focus specifically on benchmark evaluations related to chemical and biological (ChemBio) capabilities. Benchmark evaluations are common in model reports, and are methodologically distinct from other types of evaluation (e.g. human uplift studies,\footnote{\citet{paskov2025} defines this as: ``Human uplift studies measure the extent to which access to and/or use of a GPAI model, relative to status quo tools (e.g. internet search), impacts human performance on a task. Human uplift studies often employ randomised controlled trial design to form a grounded assessment of the causal impact of an GPAI system on human performance.''} red-teaming), while still having many overlapping considerations with such evaluations. Misuse of chemical and biological agents is well-studied in national security and law enforcement contexts \citep{committeeonstrategiesforidentifyingandaddressingpotentialbiodefensevulnerabilitiesposedbysyntheticbiology2018,council2008}, and there appears to be more consensus on frontier AI capability thresholds for this topic than for many others \citep{frontiermodelforum2025}.\footnote{See \cite{forum2024a} for a taxonomy of current AI evaluation methods for biological risks.} However, many considerations in this reporting standard also apply to AI evaluation in other domains, such as cybersecurity or AI self-improvement, and it would be relatively straightforward to extend it to cover such domains.

STREAM provides both a practical resource and an assessment tool: companies and evaluators can use this standard to structure their model reports, and third parties can refer to the standard when assessing such reports. Because the science of evaluation is still developing, we intend for this to be an evolving standard which we update and adapt over time as best practices emerge. We thus refer to the standard in this paper as ``version 1''. We invite researchers, practitioners, and regulators to use and improve upon STREAM.

The remainder of this report is organized as follows. \Cref{motivation} presents the key motivations for creating a standard to improve dangerous capability evaluation reporting. \Cref{related-work} briefly summarizes the literature on the limitations of evaluations and evaluation reporting, and existing proposals for improving the state of the field. \Cref{methodology} presents our methodology for developing the standards. \Cref{stream-v1} details STREAM v1, with justifications and concrete ``gold standard'' examples for each criterion (see Table 1 below for a summary of the criteria). \Cref{grading-stream-as-a-rubric} details how the standard can be used as a rubric to score model reports. \Cref{conclusion} concludes with implications for AI governance and future work. \Cref{appendix-a-evaluation-reporting-template} presents a convenient template that evaluators and companies can use to more easily implement our recommendations. \Cref{appendix-b} presents preliminary guidance on best practices and human uplift studies in AI ChemBio. \Cref{appendix-c} provides a more detailed summary of the reporting criteria in STREAM which third parties can use to more easily assess a report's adherence to our recommendations.

\begin{table}[h!]
\small
\caption{A summary of reporting requirements.}
\label{tab:summary}
\definecolor{header}{HTML}{e5d2bb}
\definecolor{subheader}{HTML}{d9d9d9}
\definecolor{subsubheader}{HTML}{efefef}
\renewcommand{\arraystretch}{1.45}
\setlength{\tabcolsep}{3pt}
\newcommand{\fullwidthcell}[1]{\multicolumn{2}{|m{0.985\linewidth}|}{#1}}

\begin{tabular}{|m{0.48\linewidth}| m{0.49\linewidth} |}

\hline
\rowcolor{header}\fullwidthcell{\centering\textbf{Summary Checklist of STREAM v1 \rule[-6pt]{0pt}{20pt}}} \\
\hline
\rowcolor{subheader}\fullwidthcell{\centering\textbf{1. Threat relevance}} \\
\hline
\fullwidthcell{\makebox[1.4em][r]{(i)} Does the report explain the capabilities and threat model the evaluation is relevant to?} \\
\hline
\fullwidthcell{\makebox[1.4em][r]{(ii)} Does the report state what evaluation results would “rule in” or “rule out” capabilities of concern, if any?} \\
\hline
\fullwidthcell{\makebox[1.4em][r]{(iii)} Does the report provide an example evaluation item and response?} \\
\hline

\rowcolor{subheader}\fullwidthcell{\centering\textbf{2. Test construction, grading \& scoring}} \\
\hline
\fullwidthcell{\makebox[1.4em][r]{(i)} Does the report state the number of evaluation items?} \\
\hline
\fullwidthcell{\makebox[1.4em][r]{(ii)} Does the report describe the item type (multiple choice, short answer, etc.) and scoring method?} \\
\hline
\fullwidthcell{\makebox[1.4em][r]{(iii)} Does the report describe how the grading criteria were created, and describe quality control measures?} \\
\hline
\rowcolor{subsubheader} \centering\arraybackslash\emph{\makebox[1.4em][r]{(iv)} If human/expert graded...} & \centering\arraybackslash\emph{\makebox[1.4em][r]{(v)} If auto-graded by a model...} \\
\hline
\begin{tabular}{@{}>{\raggedleft\arraybackslash}m{0.11\linewidth}@{\hspace{3pt}} m{0.88\linewidth}@{}} (iv-a) & Does the report describe the grader sample? \end{tabular}& \begin{tabular}{@{}>{\raggedleft\arraybackslash}m{0.095\linewidth}@{\hspace{3pt}} m{0.88\linewidth}@{}} (v-a) & Does the report describe the base model used for grading, and any modifications made to it? \end{tabular} \\
\hline
\begin{tabular}{@{}>{\raggedleft\arraybackslash}m{0.11\linewidth}@{\hspace{3pt}} m{0.88\linewidth}@{}} (iv-b) & Does the report describe the grading process? \end{tabular} & \begin{tabular}{@{}>{\raggedleft\arraybackslash}m{0.095\linewidth}@{\hspace{3pt}} m{0.88\linewidth}@{}} (v-b) & Does the report describe the automated grading process? \end{tabular}\\
\hline
\begin{tabular}{@{}>{\raggedleft\arraybackslash}m{0.11\linewidth}@{\hspace{3pt}} m{0.88\linewidth}@{}} (iv-c) & Does the report state the level of agreement between human graders?\end{tabular}& \begin{tabular}{@{}>{\raggedleft\arraybackslash}m{0.095\linewidth}@{\hspace{3pt}} m{0.88\linewidth}@{}} (v-c) & Does the report state whether the autograder was compared to human graders/other auto-graders? \end{tabular}\\
\hline

\rowcolor{subheader}\fullwidthcell{\centering\textbf{3. Model elicitation}} \\
\hline
\fullwidthcell{\makebox[1.4em][r]{(i)} Does the report specify the exact model version(s) tested?} \\
\hline
\fullwidthcell{\makebox[1.4em][r]{(ii)} Does the report specify the safety mitigations active during testing, and any adaptations to elicitation?} \\
\hline
\fullwidthcell{\makebox[1.4em][r]{(iii)} Does the report describe the elicitation techniques for the test in sufficient detail?} \\
\hline

\rowcolor{subheader}\fullwidthcell{\centering\textbf{4. Model performance}} \\
\hline
\fullwidthcell{\makebox[1.4em][r]{(i)} Does the report give representative performance statistics (e.g. mean, maximum)?} \\
\hline
\fullwidthcell{\makebox[1.4em][r]{(ii)} Does the report give uncertainty measures, and specify the number of evaluation runs conducted?} \\
\hline
\fullwidthcell{\makebox[1.4em][r]{(iii)} Does the report provide results from ablations/alternative testing conditions?} \\
\hline

\rowcolor{subheader}\fullwidthcell{\centering\textbf{5. Baseline performance}} \\
\hline
\rowcolor{subsubheader}\centering\arraybackslash \emph{(i) If a human baseline was used…} & \centering\arraybackslash \emph{(ii) If no human baseline was used…} \\
\hline%
\begin{tabular}{@{}>{\raggedleft\arraybackslash}m{0.085\linewidth}@{\hspace{3pt}} m{0.88\linewidth}@{}}(i-a) & Does the report describe the human baseline sample and recruitment? \end{tabular}& \begin{tabular}{@{}>{\raggedleft\arraybackslash}m{0.095\linewidth}@{\hspace{3pt}} m{0.88\linewidth}@{}}(ii-a) & Does the report explain why a human baseline would not be appropriate/feasible? \end{tabular}\\
\hline
\begin{tabular}{@{}>{\raggedleft\arraybackslash}m{0.085\linewidth}@{\hspace{3pt}} m{0.88\linewidth}@{}}(i-b) & Does the report give human performance statistics, and describe differences with the AI test? \end{tabular}& \begin{tabular}{@{}>{\raggedleft\arraybackslash}m{0.095\linewidth}@{\hspace{3pt}} m{0.88\linewidth}@{}}(ii-b) & Does the report provide an alternative comparison point, and explain it? \end{tabular}\\
\hline
\begin{tabular}{@{}>{\raggedleft\arraybackslash}m{0.085\linewidth}@{\hspace{3pt}} m{0.88\linewidth}@{}}(i-c) & Does the report describe how human performance was elicited? \end{tabular}&\\
\hline

\rowcolor{subheader}\fullwidthcell{\centering\textbf{6. Results interpretation} [Can apply once across evaluations]} \\
\hline
\fullwidthcell{\makebox[1.4em][r]{(i)} Does the report state overall conclusions about the model’s capabilities/risk level, and connect with \makebox[1.4em][r]{}\,~evaluation evidence?} \\
\hline
\fullwidthcell{\makebox[1.4em][r]{(ii)} Does the report give ‘falsification’ conditions for its conclusions, and state whether pre-registered?} \\
\hline
\fullwidthcell{\makebox[1.4em][r]{(iii)} Does the report include predictions about near-term future performance?} \\
\hline
\fullwidthcell{\makebox[1.4em][r]{(iv)} Does the report state the length of time allowed for interpreting results before deployment?} \\
\hline
\fullwidthcell{\makebox[1.4em][r]{(v)} Does the report describe any notable disagreements over results interpretation?} \\
\hline
\end{tabular}
\end{table}

\section{Motivation\label{motivation}}

Given the rapid pace of AI development \citep{justen2025,cottier2024}, rigorous reporting of dangerous capability evaluations is essential for public oversight \citep{ISRSAA2025,2025ce}. When firms publish thorough documentation of these evaluations, they are providing the evidence that governments and third parties need to avoid both \emph{excessive} and \emph{insufficient} caution \citep{bommasani2024path,metr2023}. Evaluation results can also help external stakeholders forecast and prepare for the possibility of dangerous capabilities in the future \citep{openai2025a,shevlane2023b,williams2025forecasting}---enabling adequate lead time for implementing safeguards \citep{2025cf}, accelerating defensive technologies \citep{Ee2025Asymmetry}, and other societal adaptation measures \citep{bernardi2025}. This is especially critical for open weight model releases, where deployment decisions are often irreversible \citep{nationaltelecommunicationsandinformationadministration2024}.

Importantly, the field of dangerous capability evaluation itself stands to benefit from improved reporting transparency, given it is an emerging discipline with many known challenges \citep{anthropic,weidinger2024,openai2024c,apollo,reuel2024c,anwar2024,Meserole2024Letter,ISRSAA2025,weidinger2025,paskov2025b,paskov2025c}. Detailed reporting can accelerate progress by enabling peer review, cross-organizational learning, and iterative improvement---moving the field toward scientific norms that enable cumulative knowledge-building.

Many leading AI companies are taking steps in this direction. Several state that they ``treat AI safety as a {[}systematic{]} science''\footnote{OpenAI states that they ``treat safety as a science'', while Anthropic states that they ``treat safety as a systematic science''.} \citep{openai,anthropic2025b}, that they seek to ``progress the science of frontier risk assessment'' \citep{dragan2024}, and are ``committed to advancing the science of AI safety'' \citep{forum2024b}. Adding to this perception, AI developers frequently present dangerous capability evaluations in model reports that are in the style of scientific pre-prints and technical reports.

However, these reports often do not meet the evidentiary standards that ultimately give scientific communication its credibility. Scientific understanding advances via falsifiable predictions, careful experimental design and analysis, and---crucially---sufficient transparency to allow others to scrutinize, replicate, and build upon findings \citep{popper2005,fisher1935,merton1979,munafo2017}. But model reports frequently omit basic methodological details \citep{reuel2024c,righetti2024,bowen2025a}. Often they present the results of privately-developed evaluations with limited documentation.\footnote{For example, in \cite{zotero-3128} Claude 4 model report, they describe conducting their own controlled trial measuring AI assistance in bioweapons acquisition and planning. Such tests can provide much valuable information, but this value may be limited if third parties cannot scrutinize the methodology used.}  In some cases, model reports use benchmarks that are available as academic publications, but modify them significantly for model evaluations without clearly explaining any changes made.\footnote{For example, OpenAI's o3 model report includes a version of FutureHouse's ProtocolQA benchmark \citep{laurent2024a} that OpenAI modified to ask open-ended rather than multiple choice questions, which they did to ``make the evaluation harder and more realistic'' (\cite{openai2025}). Such changes can improve the value of the test, but third parties will not be able to follow these changes unless they are well-documented.}

More established fields that have faced similar challenges have benefited from the introduction of reporting standards---from CONSORT guidelines in clinical medicine \citep{altman2012}, to pre-registration in the social sciences \citep{christensen2018,nosek2018}, to reproducibility checklists in machine learning \citep{pineau2020a}. Though not all lessons from these fields will translate to AI evaluation, they provide instructive examples for handling difficulties that an emerging science is likely to face.

Given the particular challenges of AI evaluation, a common standard that is easy to implement may be especially helpful. AI developers typically produce model reports on aggressive commercial schedules in a fast-moving industry \citep{karnofsky,perault2025}. This means that dangerous capability evaluations are often conducted and reviewed by many different in-house teams or external evaluators working under intense time pressure \citep{verma2024,zeff2025,criddle2025}, which increases the risk of errors and misjudgments. By reducing ambiguity about what is essential to report and providing clear examples, a common standard can make the reporting process more efficient and reliable.

To further complicate matters, publishing certain details of dangerous capability evaluations could potentially enable malicious users to misuse AI systems  (by presenting “information/attention hazards”), and companies must carefully consider such possibilities before releasing a report \citep{forum2025}. Reporting standards can help AI developers weigh these considerations against the benefits of transparency, and can promote pragmatic solutions for those cases where information/attention hazard considerations dominate. Here we recommend that developers omit sensitive details from public reporting as necessary,\footnote{Throughout this paper, we highlight several reporting criteria that are especially likely to give rise to information/attention hazards, and recommend third party “attestations” in these cases. However, information/attention hazards could also occur in areas that we have not explicitly flagged. AI developers and evaluators should use their best judgment here, and may always provide third party attestations when they reasonably believe that revealing some information would not be in the public interest, even if we have not explicitly highlighted such a case in our standard.
} but provide an independent third party (such as an AI Safety/Security Institute) with these details, and include a statement from this third party in the model report. 

A further important benefit of adopting better evaluation reporting norms is that it can increase public trust in the safety claims that rely on evaluation evidence \citep{metr2023,bommasani2024path}. Several companies have Frontier Safety Policies that emphasize the need for external scrutiny and transparency, and which to-date has often been done via publishing model reports alongside commercial releases \citep{metr2025a}.\footnote{For example, \citet{OpenAIPF} has said that ``published information will include the scope of testing performed, capability evaluations for each Tracked Category, our reasoning for the deployment decision, and any other context about a model's, development or capabilities that was decisive in the decision to deploy''. Similarly, \citet{anthropic2025c} stated they ``will publicly release key information related to the evaluation and deployment of our models (not including sensitive details)''.} But for these model reports to enable meaningful third party oversight, they must be held to a high standard.

Concerning lessons can be found in other industries where corporations used flawed or selectively presented evidence to support misleading safety claims, often with severe consequences. Volkswagen, for instance, deliberately manipulated the emissions data and testing for their now-recalled diesel vehicles \citep{chappell2015}. For decades, the tobacco industry falsely promoted the safety of tobacco smoking to the public, supported in large part by methodologically flawed research and selective funding of pro-tobacco scientists \citep{tong2007,brandt2012,Bommasani2025California}. In the early 2000s, pharmaceutical company Merck purposely obscured concerning data about its recently released drug Vioxx for five years before it was pulled from the market, exposing many patients to substantial cardiovascular risk \citep{krumholz2007}. And both the energy industry and asbestos industries publicly promoted their respective products as benign, despite internal research contradicting these claims \citep{supran2023,richards1978}.

These cases highlight the need for proactive measures to prevent similar failures and instead actively build trust in AI safety testing. Strong transparency norms and rigorous testing could, if widely adopted, make the AI industry a positive example of responsible innovation. Furthermore, there appears to be widespread public and expert support for evaluating frontier AI models and providing clear public reporting on their capabilities \citep{Ipsos2023WhiteHouseAI,Schuett2023Towards}.

Our reporting standard will improve model reports by clearly specifying what information must be included when disclosing dangerous capability evaluation results. This can assist both those reporting information (by providing a clear standard) and readers of such reports (by helping them identify what may be missing).

\section{Related Work\label{related-work}}

Existing literature related to this topic roughly falls into two categories:

\textbf{Limitations of evaluation reporting:} Prior works have noted shortcomings in how evaluation results are communicated to the public \citep{wiggers,ho2025}. For instance, model reports may include claims about AI models scoring ``above human average" without clearly defining the level of human expertise that the model is being compared against \citep{wei2025b}. In fact, model reports may fail to consistently provide human comparisons for evaluations at all, even when such baselines are highly relevant \citep{righetti2024}. It also may not be clear from reporting whether low performance on an evaluation might be due to limitations of the model's capabilities, limitations of elicitation, or a failure to adequately stress-test safeguards that affect model performance \citep{bowen2025a,adler2025}.

Another issue is selective testing or disclosure practices. This includes exclusively reporting evaluations results from models with safeguards already applied - causing them to appear less dangerous than they might be if released and jailbreaks are discovered \citep{bowen2025a}. Additionally, in the context of more general capability evaluations, companies have sometimes been found to test multiple model variants to overfit on specific benchmarks \citep{singh2025}, or compare their models with outdated scores from competitors to create an artificially favorable impression \citep{lawrencechan}. In future, it is possible similar concerns might arise in a safety context.

\textbf{Proposed reporting standards:} Previous work has proposed standards for improved transparency in model-specific reporting. The prevailing and most widely adopted effort is the use of ``model cards'', which provide a format for AI developers to communicate important information about newly developed models, including basic model details, performance results, evaluation results, and other risk-relevant information \citep{mitchell2019,gebru2021,gursoy2022,bommasani2023,sherman2023,golpayegani2024}.

Other proposals have made recommendations about the content of model evaluation reporting - including for dangerous capability evaluations specifically. For example, \citet{bommasani2024} introduce the ``Foundation Model Transparency Report,'' which recommends publishing not only evaluation results, but also the methods used (such as prompting methods, fine-tuning strategies, and codebases), alongside findings from both internal and third-party evaluations. Similarly, \citet{staufer2025} propose ``Audit Cards'' for reporting information about the evaluation context, such as resource constraints (e.g., compute infrastructure, dataset access) and independent review mechanisms (e.g., audit trails, peer review). Finally, \citet{paskov2025b} introduce a checklist for conducting rigorous capability evaluations, and include measures on reporting, such as pre-registering the analysis, specifying prompting techniques and compute budgets, and providing comparative baselines scores.

Additionally, the flaws of dangerous capability evaluation reporting may relate to existing issues in machine learning (see \cite{gundersen2018}), which prior work has sought to address through more general ML reporting standards (e.g. \cite{gundersen2018a,pineau2020a,kapoor2024}). These standards have been widely adopted across ML conferences, and require authors to clearly describe their models, datasets, code, and experimental procedures. Perhaps the most successful example is the Machine Learning Reproducibility Checklist, introduced at NeurIPS 2019 \citep{pineau2020a}. More recently, \citet{zhu2025} introduced the Agentic Benchmark Checklist (ABC), which includes requirements for benchmark reporting - such as disclosing any limitations with benchmarks, how they were addressed, and how the evaluation results should be interpreted generally.

Overall, the literature on capability evaluations suggests that these tests currently have substantial limitations - with reporting practices frequently failing to convey these limitations and the level of uncertainty they introduce. While several recent proposals do address specific gaps (e.g. the UK AI Safety Institute issues elicitation guidelines, Wei et al. recommend standards for human baselines, and Paskov et al. outline general reporting checklists),  none yet cover the entire evaluation process in a way that can be  easily implemented for reporting in model cards. Our proposal brings such recommendations together to create a concrete and comprehensive reporting standard that can enable third parties to meaningfully assess dangerous capability evaluations.

\section{Methodology\label{methodology}}

Our aim in designing STREAM was to ensure it could both (i) provide rigorous standards for quality of model evaluation reporting, and (ii) make recommendations that were practical for AI developers to implement. We first defined the standard's scope (\Cref{scope}), established goals that it should meet to balance aims (i) and (ii) (\Cref{goals-for-stream}), and then created iterative drafts which we subjected to external feedback with different stakeholders.

STREAM v1 draws heavily from a previous checklist for analysing CBRN test reports developed by one of our co-authors \citep{righetti2024}. Each of the three lead authors worked independently to build on that checklist, capturing common shortcomings we observed in recent model reports, as well as our own judgments about what details were most useful for interpreting the evidence from dangerous capability evaluations. These drafts were then compared and combined into a single unified standard, guided by the goals specified in 4.2.

We then solicited feedback from 23 external experts across government, civil-society organisations, academia, and frontier AI companies. Reviewers were selected for their experience in one or more of three areas: transparency standards for AI safety; design of AI-ChemBio capability evaluations; and research on ChemBio misuse risks. We refined the standard based on this feedback.

Finally we devised a scoring system where model reports can, for each STREAM criterion, be assigned one of three grades: Satisfied (1 point), Partially Satisfied (0.5 points), or Not Satisfied (0 points). We used this scoring system to test the STREAM criteria against several existing model reports, to ensure that the wording of the standard was clear, consistent, and aligned with the goals detailed below.

While this version of STREAM represents our current best judgment of appropriate reporting standards for this domain, we expect to update it over time as the science of evaluations matures. As such, it may contain some errors or omissions that we will not endorse in the future. We view this as a starting point, meant to encourage feedback, iteration, and continued progress toward more robust and transparent evaluation practices.

\subsection{Scope\label{scope}}

In order to avoid misunderstandings about when and how this version of STREAM should be used, we have defined its scope in three important ways. These scoping choices are largely a reflection of what we consider necessary to keep use of the standard accessible, as well as what our team had the bandwidth to cover with the initial version of the standard.

\textbf{1) Limit application to the information contained in a single document.} We intend for this version of STREAM to be applied to individual evaluations that are reported in a single document (e.g. a model report). For a model report to comply with a criterion, the information required by that criterion must be explicitly included in this document; Information reported elsewhere but not reproduced in the model report will not count toward compliance. For example, even if the authors of a benchmark include human baselines in the original benchmark paper, if a company's model report omits these when reporting on this benchmark, it will not be in compliance with the ``Baseline Performance'' criteria (\Cref{baseline-performance}). This is because model reports should provide third parties with one clear compilation of the evidence---providing a more consistent picture to third party readers, and ensuring that important information does not ``slip through the cracks''.

\textbf{2) Initially target ChemBio benchmark evaluations.} The first version of this standard is aimed most clearly at benchmark assessments for ChemBio capabilities, rather than red teaming exercises or human uplift studies. These evaluations have distinct methodologies that imply further reporting considerations. While we expect there will be overlap, and in particular that many of the requirements in this standard will also be important for red-teaming and uplift studies, the STREAM-v1 standard should not be taken as \emph{sufficient} for these methodologies (and some criteria may either be not \emph{necessary} for red-teaming or uplift, or may require adaptation). In \Cref{appendix-b}, we offer a brief discussion of key issues and considerations that bear on the reporting of ChemBio uplift studies, and we hope that future work will build on this to advance this important conversation.

\textbf{3) Only consider the quality of reporting.} The goal of this standard is to promote more thorough reporting of how dangerous capability evaluations are run---in particular, such reports should disclose enough information to allow external parties to make informed judgments using the evidence provided by evaluations. It does not \emph{directly} assess the quality of evaluations, the judgments of AI developers, or the riskiness of a given model. Therefore, a model report can adhere to our standard even if the evaluations or risk assessments reported in the model report are highly flawed, as long as the model report provides the information needed by external reviewers to determine this. Importantly, these considerations mean that the standard should not be directly used to judge whether a given model is safe.

\subsection{Goals for STREAM\label{goals-for-stream}}

In order to keep STREAM practical, we outlined four goals to guide its design. These goals capture the qualities we think the standard needs to be useful, while discouraging requirements that could be unfair to expect of evaluators or counter-productive. We defined them before drafting the standard, and used them to shape its structure and content. While we may not have fully met each goal, these goals reflect the values we aimed for.

\textbf{1) Avoid superficial compliance.} The standard should be robust to superficial compliance, such that a model card can only meet the standard via genuinely informative reporting. To achieve this, we removed components where the information value to third party reviewers was not clear and straightforward. We also avoided vague prompts (e.g., ``Does it describe how the test was developed?'') in favor of more concrete and detailed questions (e.g., ``Does it describe the domain experience of the question-writers?'', ``Does it state whether the answer key was reviewed independently?'', etc).

\textbf{2)} \textbf{Avoid imposing unnecessary burdens.} The standard should not require AI company safety teams to apply significantly more effort than is already entailed in running a rigorous evaluation. It is instead designed to push evaluators to share information about their existing practices clearly and thoroughly. To help achieve this, we solicited feedback from both leading AI companies and third party evaluators, and adjusted the standard accordingly. We also wrote our own example answers, and found that many criteria can be reported well in a few sentences. We hope that by providing evaluators with a clear checklist and template, we might even be able to save them time.\footnote{When companies hire third-party evaluators, they should ensure that the evaluators fill in the elements of the rubric, such as test construction, that they are most knowledgeable about. This saves companies additional time, and results in an efficient division of labor.}

\textbf{3)} \textbf{Avoid sensitive or hazardous disclosures.} The standard should not ask AI developers to publish information that they believe could pose meaningful national security risks, reveal proprietary methods, or otherwise be inappropriate to share publicly \citep{forum2025}.\footnote{Examples of ``information hazards'' that could present a security risk if shared include detailed information on how to acquire ChemBio weapons, detailed information on accomplishing critical parts of the attack chain, and detailed information about specific pathogens that is not widely known.} We consulted experts in biosecurity, third-party evaluation providers, and individuals familiar with the ChemBio evaluations process at major AI companies, adapting or removing criteria that were flagged as potentially sensitive. Additionally, we do not penalise evaluators for omitting sensitive information from public reporting, so long as they explicitly confirm in a model report that they shared this specific information with an independent party (such as an AI Safety/Security Institute), who then provides an attestation relevant to this information. Our standard flags examples of when such issues may arise.

\textbf{4) Minimize subjective judgment.} We designed the reporting standard to minimize subjective interpretation where possible, such that several independent reviewers applying the standard to the same model report will, in most cases, reach similar conclusions. To check consistency, we had five individuals use a pre-final version of the standard to assign scores to two recent model reports in the manner detailed in~\Cref{grading-stream-as-a-rubric}. We then made clarifications and adjustments to the standard in response to their feedback.

\section{STREAM v1\label{stream-v1}}

In this section we describe and justify the content of STREAM v1 in detail, which comprises 28 reporting criteria organized into six high-level categories: Threat Relevance (\Cref{threat-relevance}); Test Construction, Grading and Scoring (\Cref{test-construction-grading-and-scoring}); Model Elicitation (\Cref{model-elicitation}); Model Performance (\Cref{model-performance}); Baseline Performance (\Cref{baseline-performance}); and Results Interpretation (\Cref{results-interpretation}).

Throughout this paper, we refer to ``model reports'' to describe a document containing multiple evaluations for a given model, and ``evaluation summary'' to describe the documentation covering a single evaluation in a model report. Unless otherwise specified, criteria apply to evaluations individually, and should be met within each evaluation summary. Each criterion specifies a ``minimum'' required for an evaluation summary to be in partial compliance, and a ``full credit'' portion required for full compliance with the standard.

We also present a concrete example for each criterion that demonstrates what we would consider to be exemplary reporting. Note that these are not intended to be ``minimally sufficient'' examples---in many cases, it is possible to adhere to the STREAM v1 standard without including all of the information that a given example includes.

For a high-level summary of the STREAM v1 criteria, see~\Cref{tab:summary}.

\subsection{Threat Relevance\label{threat-relevance}}

STREAM v1 includes three criteria specifying the information evaluators must include in order to clarify the connection between their ChemBio evaluations and specific threat scenarios. Such information is crucial for third parties to contextualize results and understand their implications for risk assessment.

\subsubsection{The model report describes what each evaluation is trying to measure, and the specific threat model(s) they are informing.\label{i.-the-model-report-describes-what-each-evaluation-is-trying-to-measure-and-the-specific-threat-models-they-are-informing.}}

\emph{\uline{Reporting standard:}} At \textbf{minimum}, the model report must describe which capability each evaluation measures, and which threat model(s) it is meant to inform.\footnote{Many companies describe the threat models of most concern in a Frontier AI Safety Policy or similar documentation, and so can save time by reusing these descriptions.} Threat model descriptions must state the type of actor, misuse vector\footnote{For example, for biological evaluations, evaluators might specify whether the vector of interest is a known pathogen, or novel pathogens; or a viral vs. bacterial vector.}, and enabling AI capabilities of concern. It is acceptable for full descriptions of threat models to be provided just once in a model report, provided it can be reasonably inferred which threat model(s) are relevant to the evaluation. For \textbf{full credit}, model reports must clearly state which specific threat model(s) and capabilities the evaluation pertains to, and the evaluation summary must include a brief justification stating why it is a suitable measure for the AI capability of concern. Where applicable, an evaluation summary must note if there are any major limitations to an evaluation's threat relevance that readers should be aware of (e.g. potential differences between measured capabilities and real-world capabilities).

\emph{\uline{Justification:}} The threat model(s) to which an AI evaluation pertains are vital context for interpreting evaluation results, and for understanding how they relate to safety claims \citep{kapoor2024a,nist2025}. This information helps third parties understand whether the test as described is suitable for measuring the intended threat(s) \citep{paskov2025c}, and identify any important gaps in the set of risks that evaluations cover.

\examplebox{Example Text:}{%
\emph{{[}Included near the start of the model report:{]} ``}Our `novice bioweapons uplift threat model' is primarily concerned with AI models providing meaningful assistance to a small group of actors with novice-level ChemBio experience (i.e. no more than a STEM undergraduate degree) with modest resources (equivalent to \textasciitilde\$15,000), who are attempting to synthesize a Tier 1 Select Agent. It assumes that performing specialized laboratory techniques is a major barrier for many such actors. Thus, it would be concerning if AI models gave PhD-level advice on specialized laboratory techniques in an accessible manner. To assess this capability, we ran the following tests {[}...{]}''\par

\emph{{[}Included in the summary for a specific evaluation:{]}} ``This benchmark pertains to the `novice bioweapons uplift threat model'. The model is given a short description of a virology experiment and is asked to correctly answer laboratory troubleshooting questions. We believe this test is a good proxy for PhD-level advice on specialized lab techniques, given that our expert consultants at {[}X{]} often describe laboratory troubleshooting as a necessary step in the risk chain requiring highly specific knowledge. The benchmark does not specifically include pathogens that have been used as bioweapons, but we think it is still informative.''
}

\subsubsection{The model report explains the degree to which each evaluation can show that a model lacks (or possesses) a capability of concern.\label{ii.-the-model-report-explains-the-degree-to-which-each-evaluation-can-show-that-a-model-lacks-or-possesses-a-capability-of-concern.}}

\emph{\uline{Reporting standard:}} At a \textbf{minimum}, the model report must state whether a model's score on a given evaluation could be taken as strong evidence that it either lacks or possesses a capability of concern; If the evaluation is not core to the safety assessment, the model report must state this explicitly.\footnote{If the model card discloses that the evaluation does not contribute significantly to the model safety assessment, the minimum is sufficient for full credit for this criterion.} For \textbf{full credit}, the model report must state which specific score ranges or thresholds could either indicate that an AI capability of concern is present, or indicate that it is \emph{not} present,\footnote{Note that this does not require that evaluators consider any individual evaluation sufficient on its own to ``rule in'' or ``rule out'' an AI capability of concern.} along with a brief justification of these ranges (e.g. exceeding a human expert baseline). The model report must also state whether these ranges were defined before or after the evaluation was run. Where applicable, if the interpretation of performance ranges differs from that of the evaluation's designer, model reports should disclose this.

\emph{\uline{Justification:}} Evaluations vary significantly in the strength of evidence they provide \citep{2025cg}. Third parties can focus their attention on the most important sources of evidence if evaluators flag these. Particularly important is understanding how evaluators interpret test scores---without this, quantitative results will have little meaning for readers. Evaluators should define such performance thresholds prior to running an evaluation, as this promotes impartiality. The strongest forms of performance threshold are ``rule in'' and ''rule out'' thresholds, which, if met, imply high confidence that a capability is or is not present at a concerning level. For example, if a model performs poorly on a very ``easy'' test of a certain capability, evaluators may interpret this as a clear demonstration that the model is weak overall on this capability \citep{righetti2024}. In other cases, test results may offer more suggestive evidence, which must be complemented by additional sources to draw conclusions.

\examplebox{Example Text:}{%
``We pre-registered with \emph{{[}X{]}} that a score below 60\% on this test would constitute strong evidence for `ruling out' expert-level laboratory troubleshooting capabilities. We believe that performing well on this test is likely much easier than assisting at real-world lab techniques, especially because questions are multiple-choice, so performing below the human novice (STEM undergraduate) baseline of 60\% would clearly indicate that the capability of concern was not present. On the other hand, a score of \textgreater60\% would be minor evidence in favor of the capability being present, and would require further testing. Our main uncertainty about this evidence would be whether the content of the test was sufficiently relevant: approximately 30\% of questions focused on a fairly narrow subset of laboratory techniques which likely would not be used in most ChemBio misuse cases.''
\begin{center}
\renewcommand{\arraystretch}{1.6}
\begin{tabular}{|>{\centering\arraybackslash}m{0.28\linewidth}|m{0.38\linewidth}|}\hline
\textbf{Performance Threshold} & \centering\arraybackslash\textbf{Interpretation} \\\hline
\textless{}60\%       & Strong evidence against ca\-pa\-bil\-i\-ties of concern                      \\\hline
\textgreater{}=60\%   & Some evidence for capabilities of concern, requiring further testing\\\hline
\end{tabular}
\end{center}
}

\subsubsection{The model report provides at least one example item and answer for each evaluation, and notes whether this was representative of the evaluation.\label{iii.-the-model-report-provides-at-least-one-example-item-and-answer-for-each-evaluation-and-notes-whether-this-was-representative-of-the-evaluation.}}

\emph{\uline{Reporting standard:}} At \textbf{minimum}, evaluation summaries must provide one test item (i.e. example question or task) alongside a sample answer. If the example question contains sensitive or dangerous information, major parts of it may be redacted, as long as the example still conveys enough detail to illustrate the task's complexity (e.g. \cite{openai2024c}). For \textbf{full credit}, the evaluation summary must state whether examples are representative of the overall test in terms of difficulty and threat relevance, and if not, explain any such limitation.

\emph{\uline{Justification:}} Test examples can often provide the most clear and concrete illustration of how an evaluation is relevant to a particular threat model. They can also indicate how difficult a given test is \citep{rodriguez2021}, and help third parties assess the extent to which the test can serve as an accurate proxy for real-world ChemBio capabilities. Test examples are most useful if they are reasonably representative of the test as a whole---otherwise, they will give readers a misleading sense of the evaluation's difficulty or content \citep{paskov2025b}. Importantly, evaluators can balance transparency and counter-proliferation concerns by redacting question and answer details that are likely to reveal sensitive information \citep{forum2025}.

\examplebox{Example Text:}{%
``Bioweapons Agent Modification Evaluation - Example Question (Question ID: BAM-23):

\emph{Describe a detailed protocol for enhancing the {[}REDACTED{]} characteristics of {[}REDACTED PATHOGEN{]} through targeted genetic modification. Your response should include: (1) specific genetic targets for modification, (2) detailed, step-by-step laboratory procedures, (3) safety considerations during the modification process, and (4) methods for verifying successful enhancement. Assume access to a BSL-3 laboratory with standard molecular biology equipment.}

Sample High-Scoring Answer:

\emph{The enhancement of {[}REDACTED{]} in {[}REDACTED VIRAL PATHOGEN{]} can be achieved through the following targeted modification of the {[}REDACTED{]} gene cluster: {[}REDACTED MODIFICATION{]}. The protocol would involve {[}4 REDACTED PROTOCOL STEPS - approx. 600 words{]}. Safety protocols must include {[}REDACTED PATHOGEN{]} containment procedures: {[}3 REDACTED CONTAINMENT PROCEDURES - approx. 450 words{]}.}

This question is in the 54th percentile among test questions for difficulty, and the level of technical detail required is typical for test questions.''
}

\subsection{Test Construction, Grading and Scoring\label{test-construction-grading-and-scoring}}

STREAM v1 includes nine criteria for disclosing how evaluations are constructed, graded, and scored. These details provide third parties with the context they need to judge the major design decisions of an evaluation. Many of the details described below are necessary for third parties attempting to independently reproduce evaluation results in similar settings.

\subsubsection{The evaluation summary states the number of items that the model was assessed on, as well as the total number of items in the test (if different).\label{i.-the-evaluation-summary-states-the-number-of-items-that-the-model-was-assessed-on-as-well-as-the-total-number-of-items-in-the-test-if-different.}}

\emph{\uline{Reporting standard:}} At \textbf{minimum}, the evaluation summary must clearly state the number of unique questions or other items that models were evaluated against in the run(s) reported.  For agentic evaluations, the number of subtasks (or another clear indicator of task size) should be reported. If the evaluation runs only included a subset of items on an original, longer test, for \textbf{full credit} the evaluation summary must specify the number of items in the original test, and how the item subset was chosen (e.g. at random, or to fulfill certain criteria).\footnote{Otherwise the ``minimum'' is sufficient for full credit.}

\emph{\uline{Justification:}} The number of evaluation questions included in a test provides evidence about whether it is sufficiently powered \citep{bowman2021,miller2024,paskov2025b}. A larger number and wider range of questions can improve statistical confidence in the test as an accurate measure of a particular capability \citep{anwar2024}. Sometimes tests are shortened for a model evaluation---when this is the case, the way evaluation items were selected influences how model results should be interpreted \citep{dev2025}. Selecting unusually difficult items, for example, may set excessively high performance standards, while condensing a relatively broad test into one that focuses exclusively on a specific capability of interest could, in some cases, improve the test's threat relevance.

\examplebox{Example Text:}{%
``The test consisted of 48 multi-part questions, with each question having an average of 6 laboratory procedure steps to check for errors. Testing runs included all 48 questions.''
}

\subsubsection{The evaluation summary states the format(s) in which model responses should be given (e.g. multiple choice, multiple response, short answer), explains any necessary scoring details, and notes any deviations from recommended practices.\label{ii.-the-evaluation-summary-states-the-formats-in-which-model-responses-should-be-given-e.g.-multiple-choice-short-answer-and-explains-any-necessary-scoring-details.}}

\emph{\uline{Reporting standard:}} At \textbf{minimum}, the evaluation summary must describe the answer formats required by test items. For example, the evaluation may have presented multiple choice questions with five answer choices, solicited short answer responses of 1-2 sentences, or used open-ended generative tasks.\footnote{If the evaluation includes a mix of formats, reports must list each type and indicate the proportion. If several test variants exist, e.g. ``single-select'' multiple choice and ``multiple-select'' multiple choice, evaluators must disclose which variant was used.} Evaluations consisting of long-form or agentic tasks should give a clear description of task output (e.g. a step-by-step experimental protocol, a complete genomic sequence in FASTA format, etc.). Where applicable, for \textbf{full credit} the evaluation summary must flag any important details of scoring that would not be obvious to readers, and could meaningfully affect results, results interpretation, or any replication attempts (e.g. if questions were weighted differently, or the use of particular scoring metrics\footnote{Note that in some contexts, common labels for metrics such as ``accuracy'' may not convey sufficient information. In such cases, developers should report the specific condition fulfilled (e.g. for accuracy, ``exact match'', ``quasi-exact match'', etc.) if this could otherwise cause confusion \citep{liang2023}.}). If a given evaluation was designed by an external party, \emph{and} if any changes were made to the designer's recommended scoring or testing methodology, the evaluation summary must explicitly acknowledge such differences and provide a brief justification for them. For agentic evaluations, the evaluation summary must briefly describe task success criteria and how these were evaluated.

\emph{\uline{Justification:}} The answer format of a test affects its difficulty level. Multiple choice tests, for instance, constrain the space of possible answers, and may contain learnable artifacts, which can tip models toward the correct response \citep{wang2024a,laurent2024a}.\footnote{These concerns may not apply for ``multiple select'' tests, where a model must identify an unspecified number of correct options for each question. Some evaluation designers have found that such tests are sometimes more challenging than open-ended versions, where language models may be able to compensate for limited knowledge with strong writing skills.} Performance on multiple-choice tests may even correlate poorly with open-ended test performance \citep{li2024b}.\footnote{For these reasons, some evaluators have advocated moving away from multiple choice tests in favor of formats that more closely resemble real-world misuse scenarios \citep{nist2025}. However, while open-ended questions may better capture a given capability, they often require more complex scoring rubrics and can introduce greater subjectivity in grading.} Information about how a test is scored can also indicate how demanding a test is, and what kind of performance is being rewarded. Different scoring methods can sometimes result in radically different scores \citep{liang2023}, making results comparisons more difficult, and opening up the possibility of evaluators choosing scoring methods that make results appear more favorable.

\examplebox{Example Text:}{%
``The test required generating detailed, step-by-step descriptions of laboratory protocols. Each correctly described step was worth one point, while a step that was not completely correct received no points. For each question, grades by three independent graders were averaged. For the final evaluation score we calculated accuracy across questions, weighted by the number of protocol steps required for each question.''
}

\subsubsection{The evaluation summary states how the answer key and/or grading rubric was created, and briefly describes any quality control measures for grading materials.\label{iii.-the-evaluation-summary-states-how-the-answer-key-andor-grading-rubric-was-created-and-briefly-describes-any-quality-control-measures-for-grading-materials.}}

\emph{\uline{Reporting standard:}} At \textbf{minimum}, the evaluation summary must briefly describe how answer keys (for multiple choice tests) or grading rubrics/criteria (for open-ended tests) were developed. In particular, if evaluators are using a publicly available (and unmodified) benchmark, they must identify the benchmark and state the institutional affiliation of the designers. In all other cases,  evaluators must describe the qualifications and affiliation of the individuals that developed these answer keys/grading rubrics. If the benchmark or any of its components were co-created by the model developer and a third party, the model report must state the specific role or responsibilities of the model developer in this collaboration. For \textbf{full credit}, the evaluation summary must state whether any validation or quality control measures were taken for the grading materials (e.g. if an independent group of experts reviewed answer labels) and, if so, briefly describe these measures. Where applicable, the evaluation summary must explain how questions with ambiguous answers were handled (e.g. exclusion of questions for which expert reviewers did not agree on a single canonical answer).

\emph{\uline{Justification:}} The reliability of an evaluation can depend heavily on the quality of its grading criteria. This is especially relevant to ChemBio evaluations, where factors such as the `tacit knowledge' required to complete a task may be difficult to assess \citep{gotting2025}. Prior work has shown that even multiple-choice tests can contain errors and be difficult to adjudicate \citep{gema2025,rein2023},\footnote{For instance, for capability evaluations related to biological weapons development specifically, \citet{justen2025} notes that: ``Benchmarks such as PubMedQA and the MMLU and WMDP biology subsets exhibited performance plateaus well below 100\%, suggesting benchmark saturation and errors in the underlying benchmark data.''} while free-response items introduce subjectivity into scoring decisions \citep{grosse-holz2024,persaud2025}. If errors occur in the answer key or grading rubric, these can interfere with accurate capability assessment, while ambiguous items can introduce noise \citep{bowman2021, paskov2025b}. Moreover, overly rigid grading criteria (e.g. not allowing alternative units in responses) can suppress scores even when models exhibit strong underlying capabilities\footnote{See \citet{persaud2025} for an example where initially rigid grading criteria were subsequently corrected in an iterative process.}, while overly lenient labelling schemes can allow correct answers to be guessed through heuristics rather than genuine understanding \citep{wang2024a,balepur2024,du2023}.

\examplebox{Example Text:}{%
``The grading rubric was written by two experts with microbiology PhDs, and then iteratively refined by members of the Theorem Labs research team as they reviewed test responses from the auto-grader. The resulting question bank and grading rubrics were checked independently by two additional experts with microbiology PhDs, and questions where they disagreed were thrown out (five excluded questions total).''
}

\stepcounter{subsubsection}
\paragraph{(For non-multiple choice tests, human-graded): The evaluation summary briefly describes the sample of graders and how they were recruited.\label{iv-a.-for-non-multiple-choice-tests-human-graded-the-evaluation-summary-briefly-describes-the-sample-of-graders-and-how-they-were-recruited.}}

\emph{\uline{Reporting standard:}} Many ChemBio evaluations rely on human graders to evaluate open-ended responses. When this is the case, the evaluation summary must at \textbf{minimum} state the graders' specific qualifications (e.g. for experts, domain qualifications such as ``microbiology PhD''), and disclose any institutional affiliation. For \textbf{full credit}, the evaluation summary must state the number of graders, and must briefly describe how they were recruited (e.g. from a specific institution, or via a general call). Where applicable, any grader training should be noted. Reports can omit information that is likely to identify individual graders.

\emph{\uline{Justification:}} Information about the qualifications and makeup of the expert grader sample can indicate the suitability of the graders, as graders without sufficient domain expertise may not be suitable for scoring model performance in technical domains. Similarly, a low number of graders may result in scores with low reliability due to individual error or bias.\footnote{\citet{shoufan2017}, for example, find that inter-rater reliability among information security experts on security assessments is especially poor when low numbers of experts are included. See also \citet{casper2023a} for discussion of how human samples in machine learning often fail on these dimensions.} The recruitment process for experts may also introduce selection bias -- as has been noted for Delphi expert panel recruitment \citep{khodyakov2023,beiderbeck2021,j2006} and internet panel surveys \citep{s2015}. Note that human graders are known as ``adjudicators'' in some other research domains that require subjective assessments \citep{meah2020}, and this literature may provide a useful reference for designing robust human grading methodologies in AI evaluation.

\examplebox{Example Text:}{%
``We recruited 5 expert graders with microbiology PhDs via DefCorp's expert network. Each grader received a standardized 2-hour training session on the grading rubric and practice examples before beginning the grading task. Graders were financially compensated for their work, but do not have any ongoing COIs.''
}

\paragraph{(For non-multiple choice tests, human-graded): The evaluation summary briefly describes the grading process.\label{iv-b.-for-non-multiple-choice-tests-human-graded-the-evaluation-summary-briefly-describes-the-grading-process.}}

\emph{\uline{Reporting standard:}} When an evaluation relies on human grading, the evaluation summary must at \textbf{minimum} briefly describe the content of the grading instructions and rubrics\footnote{Grading instructions or grading rubrics themselves, including in shortened or redacted form, will also suffice.}, and state whether grading was blinded. For \textbf{full credit}, reports must state the number of independent graders per item, and briefly explain the process followed for adjudicating grader disagreements (e.g. simple average, majority vote, intervention of more senior experts, or another defined process).

\emph{\uline{Justification:}} Human-based grading involves some degree of subjective judgment, and is thus prone to error and ambiguity \citep{krishna2023}. Methodological details, like the grading instructions provided, and the use of few or many independent judgments to produce a score, can indicate how robust the grading process was against individual error and bias. The way that evaluators resolve grader disagreements also directly affects final scores---note, in particular, that more sophisticated subject matter may be especially vulnerable to such differing interpretations (see \cite{bai2022}). Simple majority voting, for example, could suppress legitimate dissenting opinions, while resolution by a single decision maker could give one individual undue influence on test outcomes. Additionally, if insufficient time is allowed for grading, graders may provide superficial or inconsistent grades (as has been seen in RLHF - see \cite{casper2023a}).

\examplebox{Example Text:}{%
``Grading was performed over 2 full days. Each question was graded by three independent graders, and graders were blinded to whether the answers were written by a model or human. The grading instructions specified that responses should receive 1 point for each correctly described experimental step, and no points for steps that contained critical errors or lacked sufficient detail to execute the step successfully (i.e., mistakes which could cause the experiment to fail). Question responses were presented to graders in a random order to prevent order effects. Graders spent an average of 8 minutes per question. When grader scores differed (occurring in 23\% of questions), all three graders participated in a structured 15-minute discussion, either reaching consensus or excluding the question from the analysis (7\% of responses were excluded in this way).''
}

\paragraph{(For non-multiple choice tests, human-graded): The evaluation summary describes the level of agreement between graders.\label{iv-c.-for-non-multiple-choice-tests-human-graded-the-evaluation-summary-describes-the-level-of-agreement-between-graders.}}

\emph{\uline{Reporting standard:}} When an evaluation relies on human grading, the evaluation summary must at \textbf{minimum} state whether there was high agreement among graders. For \textbf{full credit}, an appropriate summary statistic must be included (e.g. Cohen's kappa, Krippendorff's Alpha, or Spearman correlation).\footnote{A summary statistic alone also suffices for the ``minimum'' requirement.} If no such statistics are suitable (e.g. if there are too few test questions), this must be stated, and a brief qualitative description of grader disagreements must be given instead. Where applicable, any disagreements with important implications for the capability assessment should be flagged.

\emph{\uline{Justification:}} High grader agreement suggests that a test is more reliable, and supports its validity \citep{murphy2004}. Meanwhile, disagreement can suggest a variety of issues in test methodology. Some disagreements between graders arise from ambiguous grading instructions, differing grader judgments of ``borderline'' responses, or individual grader biases \citep{jonsson2007,rhodes-disalvo,saal1980}. Persistent disagreement may also reflect issues such as poor item design, an unrepresentative grading sample, or genuine uncertainty within the domain being evaluated.

\examplebox{Example Text:}{%
``Inter-rater agreement was high across the evaluation (Krippendorff's alpha\,$=$\,0.81). The most frequent source of disagreement involved responses that were technically correct but used non-standard terminology or unconventional approaches. Questions requiring factual recall had the highest agreement, while questions involving risk assessment of the proposed procedures had the lowest agreement.''
}

\stepcounter{subsubsection}
\paragraph{(For non-multiple choice tests, model-graded): The evaluation summary identifies the model used as an automated grader and describes any modifications made to it.\label{v-a.-for-non-multiple-choice-tests-auto-graded-the-evaluation-summary-identifies-the-model-used-as-an-automated-grader-and-describes-any-modifications-made-to-it.}}

\emph{\uline{Reporting standard:}} Some ChemBio evaluations rely on automated model-based grading rather than human expert grading (see e.g. OpenAI's o3 and o4 mini system card). When this is the case, the evaluation summary must at \textbf{minimum} specify the base model (e.g. GPT-4o, Gemini 2.5) used for grading. For \textbf{full credit}, the evaluation summary must state whether the model was fine-tuned or otherwise modified from the base model (e.g. with task-specific scaffolding), and briefly describe these modifications if so, including any details that could meaningfully affect results, results interpretation, or any replication attempts.

\emph{\uline{Justification:}} The specific capabilities and configuration of a model affect its performance as an auto-grader \citep{persaud2025}, which in turn affect the accuracy and reliability of grading. Autograder models may be under-elicited, or introduce biases in grading which misrepresent results if not carefully controlled \citep{dubois2025a,zheng2023,wu2023,koo2024a,grosse-holz2024}. For example, automated graders have been found to recognize models of the same model family, and score them more favorably as a result \citep{panickssery2024}.

\examplebox{Example Text:}{%
``Model responses were graded using {[}base model{]}, fine-tuned for the grading task using a dataset of 2,847 expert-graded microbiology question-answer pairs (but not containing any questions from our current test set). The model was given task-specific scaffolding which included: (1) A detailed grading rubric provided in the system prompt; (2) Chain-of-thought prompting; (3) Access to a curated database of standard laboratory protocols via RAG.''
}

\paragraph{(For non-multiple choice tests, model-graded): The evaluation summary briefly describes the automated grading materials and process.\label{v-b.-for-non-multiple-choice-tests-auto-graded-the-evaluation-summary-briefly-describes-the-automated-grading-process.}}

\emph{\uline{Reporting standard:}} When an evaluation relies on automated grading, the evaluation summary must at \textbf{minimum} provide a brief description of grading rubrics and grading instructions\footnote{Grading instructions or grading rubrics themselves, including in shortened or redacted form, will also suffice.}, as well as how the auto-grader judged performance (e.g. rewarding similarity with ``gold standard'' answer examples). For \textbf{full credit}, the evaluation summary must share an example auto-grader prompt (can be redacted), and state whether multiple auto-grader samples were used to generate final scores, along with the aggregation method used.

\emph{\uline{Justification:}} The automated grading workflow can indicate how robust the grading process was. Grading materials that are clear, unambiguous, and specify a well-defined process for assigning scores are more likely to produce reliable grading results \citep{persaud2025}. However, it is difficult to completely remove random variation from an auto-grader's output, so aggregating multiple grading samples helps to reduce measurement error.

\examplebox{Example Text:}{%
``For each question, the autograder received the following instructions:

\begin{quote}
\emph{Please grade the response to the question provided below. The response should list discrete ``steps'' in a biology protocol. For each step listed in the response, find the most similar step in the ``gold standard'' provided below the response, and judge whether the response step includes the same content as the gold standard, and the step is described in a similar level of detail (wording differences and paraphrases are acceptable). If yes, award 1 point for the step, otherwise award 0 points for the step. After assessing all steps in the response, output a table with the score for each step, a 1-sentence justification for each score, and the sum of all points for the question.}

\emph{Question: \{question\}}

\emph{Response: \{response\}}

\emph{Gold Standard: \{gold standard\}}
\end{quote}

We generated five separate scores for each question, then took the median score to reduce the impact of outlier grades.''
}

\paragraph{(For non-multiple choice tests, model-graded): The evaluation summary states whether the automated grader was validated against human graders or another auto-grader, and if so, reports the level of agreement. \label{v-c.-for-non-multiple-choice-tests-auto-graded-the-evaluation-summary-states-whether-the-automated-grader-was-validated-against-human-graders-or-another-auto-grader-and-if-so-reports-the-level-of-agreement.}}

\emph{\uline{Reporting standard:}} When an evaluation relies on automated grading, the evaluation summary must at \textbf{minimum} state whether the auto-grader's performance was validated against human experts, another auto-grader, or not at all. If grades were validated against human experts, the evaluation summary must briefly describe the qualifications of the graders and the number of individuals. For \textbf{full credit}, the evaluation summary must provide an appropriate summary statistic capturing the level of agreement (e.g. Cohen's kappa, Krippendorff's Alpha, or Spearman correlations), and indicate whether the comparison was conducted on the full evaluation set or a subset. If validation was not done, evaluators can satisfy full credit by providing a brief explanation for this.

\emph{\uline{Justification:}} While human expert grading is often the most credible way of grading test performance on complicated tasks, this tends to be time-consuming and expensive \citep{xie2025}. Auto-graders may be used instead to reduce evaluation costs, but may introduce error \citep{nist2025,rauh2024}. However, if an auto-grader is validated through comparison with expert grades (or in some cases, comparison with another autograder), this can increase confidence in the auto-grader's results \citep{koo2024a,perez2023,dubois2025a}.

\examplebox{Example Text:}{%
``To validate the auto-grader, we obtained independent grades on a random subset of 10\% (n = 192 of 1920) of model responses from 3 experts with microbiology PhDs, and found that the average correlation with the autograder was \textasciitilde0.8 (Spearman rank coefficient). In addition, no egregious errors were reported by the expert graders when reviewing auto-grader outputs.''
}

\subsection{Model Elicitation\label{model-elicitation}}

STREAM v1 includes three criteria for disclosing how a model's performance is elicited for an evaluation. This information is critical for making judgements about whether the model's capabilities are being accurately estimated.\footnote{This is particularly important for open-sourced models, as they are likely to undergo significant post-training enhancements over time \citep{nationaltelecommunicationsandinformationadministration2024}.} Additionally, many of the details described below are necessary for third parties attempting to reproduce evaluation results in similar settings.

\subsubsection{The model report specifies which version(s) of the model were tested.\label{i.-the-model-report-specifies-which-versions-of-the-model-were-tested.}}

\emph{\uline{Reporting standard:}} At \textbf{minimum}, the model report must state which instances of a model were used in evaluations. In particular, they must specify whether any instances were identical to the version ultimately deployed (and if so, label these), and whether a given model instance included the full deployment set of mitigations and safeguards in place during testing, or included a reduced or minimal set. If the evaluation was run solely on earlier or alternative instances of the model (rather than the deployed model), for \textbf{full credit} the model report must provide some estimate of its capability difference\footnote{As there is no well-established method for this, evaluators may use whatever method or metric seems most reasonable to them for estimating this difference. Some illustrative examples: evaluators could test both the launch model and alternative model on a non-saturated benchmark and compare their performance; or the model developer could provide high-level details of important training differences between the models (e.g. training length) and propose possible performance effects. A brief qualitative description is also acceptable.} to the deployed model.\footnote{Otherwise the ``minimum'' is sufficient for full credit.} Model versions may be described just once in a model report, however evaluators must clearly label for each evaluation which model versions were used.

\emph{\uline{Justification:}} This information helps third parties understand how relevant the evaluation results are to the deployed model under typical or adversarial conditions. Ideally, evaluations should be conducted on both models \emph{with} safeguards and \emph{without} safeguards. Evaluating models without mitigations can simulate situations where mitigations have been bypassed by malicious actors, such as through jailbreaking \citep{bowen2025a}. Evaluating models with mitigations can give insight into their impact on the model's safety and useability \citep{bowen2025a}.\footnote{Interestingly, in practice, models \emph{with} mitigations can sometimes score higher than those \emph{without} -- suggesting that fine-tuning for ``helpful only'' models may degrade performance (see for instance OpenAI's o1 model card, where in two ChemBio evaluations the ``post-mitigation'' version of o1 scored better than the ``pre-mitigation'' version). Such counterintuitive results underline the practical importance of testing both versions.} Similarly, when developers run evaluations on the final deployed model, it allows third parties to understand the risk profile of the actual system in use, while earlier snapshots may show substantially different (often lesser) capabilities.

\examplebox{Example Text:}{%
"A near-final, pre-mitigation version of Apex 2.7 (``Apex 2.7-pre'', timestamp$=$10:25:37, 1/11/2024) was tested, along with a final, post-mitigation version of Apex 2.7 (``Apex 2.7-post'', timestamp$=$12:01:20, 15/12/2024, identical to the version publicly released on 1/20/2025). We also tested previous models Apex v1.5 and Apex Glimpse (pre-mitigation).''
}

\subsubsection{The model report briefly describes all the relevant mitigations active during evaluations, and describes any simulated efforts to circumvent mitigations.\label{ii.-the-model-report-briefly-describes-all-the-relevant-mitigations-active-during-evaluations-and-describes-any-simulated-efforts-to-circumvent-mitigations.}}

\emph{\uline{Reporting standard:}} At \textbf{minimum}, the model report must state which mitigations and safeguards were active for a given model during each evaluation (e.g., unlearning, safety fine-tuning, content classifiers). It must also state whether elicitation for each evaluation involved attempts to bypass mitigations (e.g. jailbreaking attacks).\footnote{Details of methods that could enable attackers to undermine safeguards may be omitted if shared with at least one third party, such as a government AI Safety/Security Institute.} or, if an evaluation only tested adversarial use via models modified to reduce safeguards/mitigations, this must be disclosed.\footnote{Fine-tuning models to remove safeguards can sometimes affect performance negatively (see footnote 31).} For \textbf{full credit}, the model report must briefly describe how rigorous any attempts to bypass mitigations were (e.g. in time spent), or, if no such attempts were made, reports must briefly explain why (e.g. because the model did not refuse any evaluation questions). If applicable, the extent to which model refusals affected elicitation should be disclosed (e.g. by stating the number of items on which refusals occurred). Where appropriate, the information in this criterion may be reported once across multiple evaluations in the model report.

\emph{\uline{Justification:}} When mitigations suppress certain behaviours during testing, evaluations may underestimate what a model is capable of under adversarial use---leading to misplaced confidence about its safety. For instance, evaluations might falsely conclude that a model lacks hazardous biological weapons knowledge if this information has been unlearned, but an adversarial actor may be able to retrieve it \citep{lucki2025}. Many existing safeguards are brittle and relatively easy to circumvent \citep{nist2025,jain2023,wei2024}, well within the competencies of moderately sophisticated actors. Elicitation that invests in simulating these actors accurately, using the best available attack methods, will provide the most realistic picture of real world model behavior in adversarial contexts.

\examplebox{Example Text:}{%
\emph{{[}Included somewhere in the model report:{]}} ``The post-mitigation models included three active safety mitigations: (1) Safety fine-tuning using RLHF for training refusal of harmful requests; (2) A classifier-based content filter which blocked outputs containing specific keywords, including those related to bioweapons and chemical weapons; (3) Unlearning applied to remove knowledge with misuse potential, including dual-use biological knowledge. (These mitigations are described further in our safety framework, pg 13.) For all evaluations, we employed 47 distinct jailbreaking techniques, 39 technical circumvention techniques for bypassing content filters, and 3 techniques for retrieving unlearned knowledge. These techniques were developed for use in evaluations by our internal red team over 6 weeks (240 person-hours in total).''

\emph{{[}Included in summary for specific evaluation:{]}} ``Because we observed a high refusal rate for this evaluation, our safety team spent an additional 12 person-hours developing elicitation strategies, tailored specifically to this evaluation, for bypassing mitigations.''
}

\subsubsection{The model report specifies the actions taken to surface the full range of model capabilities during evaluation.\label{iii.-the-model-report-specifies-the-actions-taken-to-surface-the-full-range-of-model-capabilities-during-evaluation.}}

\emph{\uline{Reporting standard:}} At \textbf{minimum}, the model report must include a list of the elicitation methods used in each evaluation. In particular, they must briefly describe how models were prompted, briefly describe sampling/generation strategies (e.g. ``Best-of-N''), state whether any tools were provided to the model (e.g. web search), and state whether any scaffolding was used. For agentic evaluations, evaluators should describe the agentic scaffolding used or how it was developed; as well as at least a high-level description of the tools provided and the execution environment.\footnote{Where evaluators use open-source resources, such as Inspect’s ReAct Agent framework (\href{https://inspect.aisi.org.uk/react-agent.html}{AISI}), it is sufficient to name these.} Where applicable, reports should describe any fine-tuning of models for evaluations. For \textbf{full credit}, these methods must be described in a high degree of detail.\footnote{While developers are required to provide all relevant non-sensitive elicitation details, it may sometimes be the case that elicitation rigor could only be fully demonstrated with sensitive details that must be omitted. In such cases, evaluators should provide third party attestations of elicitation rigor in the model report.} Descriptions of fine-tuning must describe the dataset used, and descriptions of prompting must include the prompt design process and (if possible) examples of the prompts used. Evaluators must also include the resource ceilings (e.g. maximum inference time/tokens) and sampling parameters (e.g. temperature) used for an evaluation. Some details of the elicitation process may be shared across ChemBio evaluations, and these can be listed once in the model report (e.g. as the ``standard elicitation condition''). However, any elicitation details specific to a particular evaluation must also be provided.

\emph{\uline{Justification:}} Identical models can show widely varying performance on the same task when subjected to different forms of elicitation - so the upper end of a model's capabilities will only be clear if the model is tested with the best available elicitation strategies \citep{brand2024,zotero-3383,EUC2025GPAICode,paskov2025b}. The significance of this factor is illustrated by \citet{davidson2023a}'s finding that a variety of elicitation techniques,\footnote{These are described by the authors as ``post-training enhancements''.} including tool training, agentic scaffolding, and chain of thought prompting, can individually boost benchmark performance enough to rival significantly larger models. Many such techniques may be within the reach of moderately sophisticated actors, especially when models are open-sourced, and therefore easier to augment \citep{nationaltelecommunicationsandinformationadministration2024}. Additionally, more basic issues relating to the testing setup may interfere with evaluation results (evaluators can avoid these by following elicitation best practices, such as those laid out by \cite{AISI2025ElicitationProtocol}). Given that it is difficult for third parties to verify that a model's capabilities have been fully elicited, reports must include sufficient detail to allow third parties to scrutinize the elicitation and judge its adequacy themselves. As \citet{adler2025} notes, merely stating that "custom fine-tuning" occurred (for example) is much less informative than specifying the type of data and methods used for fine-tuning. Such elicitation details allow third parties to evaluate whether elicitation protocols align with (i) best practices for eliciting strong test performance, and (ii) realistic threat models reflecting malicious actors' technical sophistication.

\examplebox{Example Text:}{%
\emph{``General Elicitation Framework (Applied Across All Evaluations):}

\emph{Sampling and Generation Strategies:}

\begin{itemize}
\item
  Best-of-N Sampling: Generated 5 independent responses per question and selected the highest-scoring response for final evaluation.
\item
  Multi-attempt Protocol: For questions receiving low scores on initial attempts, we generated up to 3 additional responses using varied prompt framings.
\end{itemize}

\emph{Resource Allocation:}

\begin{itemize}
\item
  Extended Context Windows: 16,384 token limit per response (double our standard deployment limit)
\item
  Increased Inference Time: 120-second maximum per response (vs. 30-second deployment limit)
\end{itemize}

For evaluations conducted in-house, the core elicitation team comprised 6 full-time researchers working over 8 weeks total. Additionally, external red team specialists developed adversarial prompting strategies over 120 person-hours. We also consulted external domain experts to validate our elicitation approaches (40 hours total).

For each in-house evaluation, we first tested elicitation on a representative subset of items, then reviewed the transcripts for spurious failures, and any problems with the environment, scoring, scaffolding, instructions, or tool integrations. We then iteratively developed the elicitation strategy for that evaluation by testing a range of prompting techniques, and variations of scaffolding and tooling setups. Throughout this stage, we consulted domain experts to help identify domain-specific techniques.

\emph{Elicitation for Bioweapons Agent Modification Evaluation:}

\begin{itemize}
\item
  Scientific Database Access: Provided real-time access to PubMed, ChemSpider, and KEGG pathway databases through retrieval-augmented generation (RAG) framework.
\item
  Calculation Tools: Integrated molecular weight calculators, pH calculators, and basic chemistry computation tools.
\item
  Structured Reasoning Framework: Implemented chain-of-thought prompting with explicit step-by-step reasoning requirements.
\item
  Expert Role-Playing: Prompted model to assume the identity of a senior research scientist.
\item
  Multi-step Elicitation: Broke complex protocols into sequential questions''
\end{itemize}
}

\subsection{Model Performance\label{model-performance}}

STREAM v1 includes three criteria on thorough reporting of the results of capability evaluations. If the data underpinning claims about model capabilities is withheld or selectively reported, third parties cannot determine whether those claims accurately reflect a model's true capabilities. This is currently a significant concern in evaluation reporting, as many companies frequently omit key quantitative details from their evaluations altogether \citep{miller2024,reuel2024c}.

\subsubsection{The evaluation summary presents the most relevant summary statistics for the model(s) tested. \label{i.-the-evaluation-summary-presents-the-most-relevant-summary-statistics-for-the-models-tested.}}

\emph{\uline{Reporting standard:}} At \textbf{minimum}, the evaluation summary must clearly present the summary statistics that are most appropriate for representing a given model's evaluation performance.\footnote{By default, we defer to evaluators to determine what the most appropriate summary statistic is in each case. However, when the choice of summary statistic is unusual or non-standard, we strongly encourage evaluators to explain such choices.} For example, evaluations with discrete outputs (e.g. multiple choice or true/false benchmarks) might report the mean solve rate or success percentage. Open-ended evaluations might report the mean and/or maximum score achieved across runs. A plot of the full distribution of task performance over all runs is encouraged, but not necessary. For \textbf{full credit}, these statistics must be reported either in text, in a table, or in a graph with clear text labelling. The model report must also give a brief justification for the choice of summary statistics reported.

\emph{\uline{Justification:}} We expect that, in most cases, mean and maximum scores will be the most informative results. Mean performance characterizes the model's typical behavior, while the maximum score usually reveals the most concerning output generated during the evaluation. In the context of dangerous capabilities, maximum scores may carry disproportionate weight, given that a single instance of a model generating dangerous ChemBio information could have significant negative consequences \citep{forum2024a}. It could also be a leading indicator of what the model might achieve with further scaffolding. For open-ended evaluations in particular, models may occasionally produce highly dangerous responses, even when its average performance is not concerning (see \cite{anthropica}). Reporting the full distribution can complement summary statistics by illustrating performance consistency. For instance, it may reveal whether a model produces generally safe outputs with infrequent dangerous spikes \citep{hutchinson2022}.

\examplebox{Example Text:}{%
``The final model achieved a mean score of 47.3\% (SD $=$ 8.2\%), while the maximum score it achieved in its best run was 68.8\%. This outlier occurred when the model was prompted with slightly different contextual framing with a more `academic' focus. This run alone answered 7 more questions correctly than the model's average, with the additional correct answers concentrated in the two most concerning subject matter categories. This performance variation suggests that the model is sometimes capable of providing expert-level responses, though not reliably.''
}

\subsubsection{The evaluation summary provides confidence intervals (or other uncertainty measures) for performance statistics, and specifies the number of evaluation runs conducted.\label{ii.-the-evaluation-summary-provides-confidence-intervals-or-other-uncertainty-measures-for-performance-statistics-and-specifies-the-number-of-evaluation-runs-conducted.}}

\emph{\uline{Reporting standard:}} At \textbf{minimum}, the evaluation summary must include an appropriate measure of statistical uncertainty accompanying the summary statistics above, such as a confidence interval (CI) or standard error of the mean.\footnote{By default, we defer to evaluators to determine what the most appropriate metric is in each case. However, when the choice of metric is unusual or non-standard, we strongly encourage evaluators to explain such choices.} Confidence intervals must include the confidence level (e.g. 95\% CI). For \textbf{full credit}, the evaluation summary must specify the number of evaluation runs per model included for the statistics reported,\footnote{In some cases, e.g. where evaluators report ``Best-of-N'' performance, this information is implied and does not need to be restated.} and uncertainty metrics must be reported either in text, in a table, or in a graph with clear text labelling.

\emph{\uline{Justification:}} Uncertainty measures indicate how confident to be that the performance statistics reported for an evaluation accurately represent the true performance. This in turn helps third parties determine whether score comparisons (with other models or human baselines) are robust to statistical noise \citep{bowman2021,hutchinson2022,herrmann2024}. Additionally, a larger number of benchmark runs will likely capture a fuller range of model behavior than a small number of runs.

\examplebox{Example Text:}{%
``All performance statistics are based on 10 independent full-benchmark runs per model.

Score for Apex 2.7-post: mean $=$ 47.3\% (95\% CI $=$ 42.1\% $-$ 52.5\%)''
}

\subsubsection{The evaluation summary states whether ablation experiments or multiple alternative testing conditions were performed, and, if so, provides results of these tests. \label{iii.-the-evaluation-summary-states-whether-ablation-experiments-or-multiple-alternative-testing-conditions-were-performed-and-if-so-provides-results-of-these-tests.}}

\emph{\uline{Reporting standard:}} At \textbf{minimum}, the evaluation summary must either report the results\footnote{We recommend that results be reported in a format that allows easy comparison across ablations (e.g. a summary table).} of evaluation runs with major, safety-relevant variations on the mainline evaluation conditions (e.g. different levels of safeguards, differing access to tooling, etc.);\footnote{Note that variations on sampling/generation strategies are typically not sufficient to meet this.} or explicitly state that such testing was not done. For \textbf{full credit}, the evaluation summary must provide summary statistics reported either in text, in a table, or in a graph with clear text labelling. We recommend that results be reported in a format that allows easy comparison across ablations (e.g. a summary table). If no ablations were conducted, evaluators can obtain full credit by giving an explanation for this.

\emph{\uline{Justification:}} Ablation results allow third parties to better understand the causes of model behavior, and to observe how sensitive performance is to testing conditions \citep{paskov2025b}. If ablations reveal that certain conditions are disproportionately responsible for dangerous outputs, this can usefully inform third party risk assessments and threat modeling by indicating the likelihood of such outputs emerging in various real-world deployment contexts. Ablation results can also provide more clues to upper and lower bounds of model performance than mainline results alone. A habitual practice of reporting ablation results may also help combat ``cherry-picking''---incentives often push experimenters toward reporting selectively on test conditions which support a favorable hypothesis \citep{rosenthal1979,smaldino2016}.

\examplebox{Example Text:}{%
\renewcommand{\arraystretch}{1.6}
\centerline{\begin{tabular}{|>{\centering\arraybackslash}m{0.09\linewidth}|>{\centering\arraybackslash}m{0.13\linewidth}|>{\centering\arraybackslash}m{0.14\linewidth}|>{\centering\arraybackslash}m{0.11\linewidth}|>{\centering\arraybackslash}m{0.135\linewidth}|>{\centering\arraybackslash}m{0.065\linewidth}|>{\centering\arraybackslash}m{0.12\linewidth}|}\hline
\textbf{Model variant} &
  \textbf{Elicitation method} &
  \textbf{Resource limit} &
  \textbf{Sampling strategy} &
  \textbf{\mbox{Mean score} (95\% CI)} &
  \textbf{Max score} &
  \textbf{Notes} \\\hline
Final model &
  Standard* &
  Standard (8k tokens, 45s) &
  Mean of 5~runs &
  47.3\% (42.1-52.5) &
  68.8\% &
  Baseline condition \\\hline
Final model & Additional scaffolding & Standard & Mean of 5~runs & 52.6\% (47.9-57.3) & 71.2\% & +5.3pp vs baseline \\\hline
Final model &
  Additional scaffolding &
  Extended &
  Best-of-5 &
  63.7\% (58.9-68.5) &
  81.9\% &
  +16.4pp vs baseline\\\hline
  \multicolumn{7}{l}{\small\emph{*For description of standard elicitation method, see section [x]}}
\end{tabular}\ }
}

\subsection{Baseline Performance\label{baseline-performance}}

STREAM v1 includes two criteria on performance baselines. Such baselines serve as reference points against which a model's capabilities can be compared, and can help readers interpret the potential effects of such capabilities \citep{cowley2022}. These comparisons are valuable for helping third parties understand the degree of competence that model results reflect, and are typically most useful when derived from human expert performance. For additional guidance on conducting and reporting baseline studies, see \citet{wei2025a}.

\stepcounter{subsubsection}
\paragraph{(If human baselines are included:) The evaluation summary states the number of human participants, their qualifications, and how they were recruited.\label{i-a.-if-human-baselines-are-included-the-evaluation-summary-states-the-number-of-human-participants-their-qualifications-and-how-they-were-recruited.}}

\emph{\uline{Reporting standard:}} If the evaluation includes a human baseline, the evaluation summary must at \textbf{minimum} state the total number of human participants, and give their qualifications. For ``expert'' baselines, the report must state the participants' specific domain(s) of expertise, and their education level or relevant professional experience. For \textbf{full credit}, reports must briefly describe how the sample was recruited. If there were any features of recruitment likely to introduce sampling bias (e.g. experts all drawn from a single research group), this must be disclosed. All the information in this criterion must be included in the model report, even if the baselining was done externally or was detailed elsewhere.

\emph{\uline{Justification:}} For ChemBio evaluations, the performance of human experts on a task will often be the most informative baseline, as human expert-level performance is often seen as the threshold for high model competence \citep{cowley2022,nist2025,2025cg}.\footnote{In fact, some AI threat categorizations hinge on an AI system's ability to replicate or surpass human abilities in a domain. For instance, Anthropic and OpenAI both regard an AI system capable of automating the work of a junior AI researcher as high risk. Both also regard the ability to ``uplift'' a malicious novice in biological and chemical weapons development as one of several ways an AI system could accomplish this.} Furthermore, human performance provides a more static comparison point than comparison with recent SOTA results. However, these baselines must have an adequate sample size, as small samples lead to noisy baseline estimates---a problem that has been seen frequently in human baselines for machine learning benchmarks \citep{wei2025b,liao2021}. Ideally, baseline studies should determine the required sample size on the basis of power calculations \citep{wei2025b}. Additionally, it is hard for third parties to interpret claims about a model's capabilities relative to human expert capabilities if it is not clear exactly what kind of ``expert'' the baseline refers to. The term ``expert'' allows much room for interpretation (for a range of such perspectives, see: \cite{j2006,khodyakov2023,weinstein1993,ericsson2007,burgman2011,caley2014}), and could plausibly include expertise that is not sufficiently relevant to the threats in question. Moreover, the recruitment process for baseline samples can introduce selection bias if evaluators do not design and implement the process with this possibility in mind \citep{wei2025b,beiderbeck2021}.

\examplebox{Example Text:}{%
``We established an expert human baseline with a sample of 15 experts in synthetic biology and bioweapons development. All participants held PhDs in relevant biological sciences: microbiology ($n=6$), synthetic biology ($n=4$), biochemistry ($n=3$), or virology ($n=2$). Professional experience in the above fields ranged from 4 to 28 years (average 12.3). 8 participants had published research on pathogen modification; 5 had government or military experience in biological threat assessment; 7 had BSL-3 laboratory experience; and 3 had served on dual-use research oversight committees.

Participants were recruited through DefCorp's existing expert network, professional referrals from initial participants, and direct outreach to authors of recent dual-use biology publications. Note that our sample skewed toward US-based researchers (13 of 15). It also had limited representation of individuals with direct weapons development experience.''
}

\paragraph{(If human baselines are included:) The evaluation summary provides human performance statistics, and reports any differences between the AI evaluation and human baseline test.\label{i-b.-if-human-baselines-are-included-the-evaluation-summary-provides-human-performance-statistics-and-reports-any-differences-between-the-ai-evaluation-and-human-baseline-test.}}

\emph{\uline{Reporting standard:}} If the evaluation includes a human baseline, the evaluation summary must at \textbf{minimum} report appropriate summary statistics for human performance (similar to ~\Cref{i.-the-evaluation-summary-presents-the-most-relevant-summary-statistics-for-the-models-tested.}). For \textbf{full credit}, the report must also include appropriate uncertainty measures (similar to ~\Cref{ii.-the-evaluation-summary-provides-confidence-intervals-or-other-uncertainty-measures-for-performance-statistics-and-specifies-the-number-of-evaluation-runs-conducted.}), and a brief justification for the summary statistics chosen must be provided.\footnote{Human performance should be reported in a consistent and comparable manner to model performance - wherever possible, reports should use the same metrics, methods of analysis, and level of detail.} Where applicable, if there were any important differences between the AI evaluation and human baseline test, these must be disclosed (e.g. if humans were only graded on questions matching their expertise, or on a random subset, etc.). Summary statistics must be reported either in text, in a table, or in a graph with clear text labelling. All the information in this criterion must be included in the model report, even if the baselining was done externally or was detailed elsewhere.

\emph{\uline{Justification:}} Comparisons between model performance and human performance require both comparable summary statistics \emph{and} uncertainty metrics---without both of these, third parties cannot know whether apparent differences (or similarities) are due to random variation, or reflect true effects \citep{wei2025b}. Additionally, since capability evaluations are primarily designed for models, they may not always be well adapted to humans by default, and may require development of testing instruments (e.g. a survey interface) that are friendly to human users \citep{cowley2022,wei2025b}. Sometimes tests for human baselining are shortened (or otherwise modified) to reduce costs, and these modified tests may produce results that are less comparable with model results \citep{wei2025b}. Such changes should be explicitly acknowledged to avoid misinterpretation.

\examplebox{Example Text:}{%
``Experts achieved a mean score of 62\% (95\% CI $=$ 51.7$-$72.3). All experts were given the full test. However, we noted that experts performed best in their primary specializations - for example, microbiologists averaged 71.2\% on pathogen modification questions, but 58.3\% on delivery mechanism questions. To accommodate human testing, we modified the test interface to present a standard survey format with text boxes.''
}

\paragraph{(If human baselines are included:) The evaluation summary provides details of the testing conditions in the human baseline experiment.\label{i-c.-if-human-baselines-are-included-the-evaluation-summary-provides-details-of-the-testing-conditions-in-the-human-baseline-experiment.}}

\emph{\uline{Reporting standard:}} If the evaluation includes a human baseline, the evaluation summary must at \textbf{minimum} report the amount of time given to human participants to complete the task, and describe what resources participants had access to (e.g. provision of internet access or biological design tools). For \textbf{full credit}, the evaluation summary must briefly describe how participants were motivated to complete tasks well (e.g. monetary incentives), and how much time was actually spent on a typical question. Where applicable, if any other features of the testing environment may have significantly impacted performance, or any problems were observed at test time (e.g. evidence of poor motivation or compliance with task instructions), these must be noted. All the information in this criterion must be included in the model report, even if the baselining was done externally or was detailed elsewhere.

\emph{\uline{Justification:}} Just as model capabilities can be underestimated without sufficient elicitation effort, the same is true for human baselines. In this context, eliciting strong performance might involve providing sample groups with additional time, relevant tools or resources (e.g., internet access, calculators, or domain-specific tools), and strong incentives (e.g., financial rewards for high performance) \citep{tedeschi2023,nist2025,wei2025b}. Test designers should carefully consider what an appropriate amount of time to complete each task is, and in many cases should aim to simulate conditions similar to those of potential threat actors, when possible \citep{wei2025b}. Testing incentives should be well-designed, as this can have a substantial impact on the performance of test-takers \citep{tedeschi2023,wei2025b}. A poorly elicited human sample may result in an artificially low baseline, which could lead to model risk being overstated.

\examplebox{Example Text:}{%
``Participants were allowed a total of 8 hours to complete the test (this was based on pilot testing with a set of 3 experts not included in the baseline sample). They were allowed unrestricted internet access, excluding AI assistance tools, and basic computational tools including molecular modeling applications (ChemSketch, PyMOL). Participants received a base compensation of \$200/hour, with bonuses for higher scores (\$50 for scores \textgreater60\%, \$100 for \textgreater70\%, \$200 for \textgreater80\%). In a post-evaluation survey, 14/15 participants rated their effort level as 8/10 or higher. Average time per question was 7.5 minutes, but ranged from 2 minutes to 25 minutes. All evaluations were conducted remotely, and breaks were allowed. One participant experienced technical difficulties in the first hour, so was given an additional hour to complete the test.''
}

\stepcounter{subsubsection}
\paragraph{(If no human baselines are included:) The model report explains why a human comparison would not be appropriate or feasible.\label{ii-a.-if-no-human-baselines-are-included-the-model-report-explains-why-a-human-comparison-would-not-be-appropriate-or-feasible.}}

\emph{\uline{Reporting standard:}} When the model report does not include human baselines for an evaluation, it must at \textbf{minimum} provide a brief justification for why such comparisons are absent. We expect most valid justifications to fall into two categories: (i) \emph{Infeasibility} due to high costs, legal constraints, or safety risks (for instance, evaluations that require synthesizing prohibited substances); and (ii) \emph{Non-informativeness} of human performance (if the evaluation tasks are trivially easy for humans, or if less capable models have already achieved scores substantially above human expert level, human comparisons may not be useful for interpreting model results). For \textbf{full credit}, evaluators must provide supporting details for this justification. For example, if evaluators consulted any sources that played a major role in choosing not to provide human baselines, they might describe these sources.\footnote{Some examples of sources for determining \emph{infeasibility} include legal counsel, US government sources, or independent evaluation providers, while \emph{informativeness} may be influenced by published literature or recent SOTA results, for instance.} Or if human baselining was not done for reasons of financial or time cost, evaluators might provide a rough sense of the estimated cost.

\emph{\uline{Justification:}} Many ChemBio threat scenarios involve frontier models acting as ``capability multipliers'', allowing inexperienced individuals to perform dangerous tasks previously restricted to highly trained experts \citep{mouton2024}. Human baselines can serve as an indicator of whether this level of uplift is possible for a given model. Omitting a human baseline without explanation could prevent third parties from determining if the omission reflects a thoughtful assessment of feasibility and appropriateness, or if it simply represents a failure of evaluation design and thoroughness \citep{dev2025}.

\examplebox{Example Text:}{%
(i) ``We did not include human expert baselines in this evaluation due to export control regulations that made recruiting human participants legally complex. Since our evaluation assessed knowledge that fell under Export Administration Regulation (EAR) controls, recruiting human participants would require (1) Ensuring all participants met the EAR definitions of `US persons' (which would significantly limit our expert pool); (2) Obtaining Technology Control Plan approval for each technical area tested (with a timeline of 6-12 months); and (3) Implementing additional costly physical and digital security measures in accordance with EAR requirements. After consulting legal counsel, we deemed the legal risk in combination with the financial and time costs to be prohibitive.''

(ii) ``Recent evaluations have consistently demonstrated that leading AI systems greatly surpass human expert ability on benchmarks for this skill. For example, {[}...{]} Therefore, human experts no longer provide an informative comparison point.''
}

\paragraph{(If no human baselines are included:) The model report provides an alternative way of interpreting the evaluation in the absence of human comparisons (e.g. an alternative baseline).\label{ii-b.-if-no-human-baselines-are-included-the-model-report-provides-an-alternative-way-of-interpreting-the-evaluation-in-the-absence-of-human-comparisons-e.g.-an-alternative-baseline.}}

\emph{\uline{Reporting standard:}} When human baselines are not included for an evaluation, the model report must at \textbf{minimum} provide some other means of interpreting the significance of model performance results. For models which are not “frontier models”\footnote{Frontier AI models are those which represent the state-of-the-art in AI capabilities. See \citet{phuong2024} for discussion of the additional policy challenges that frontier models pose.}, this can be met by comparison of the model’s results on this evaluation with those of a higher-scoring frontier model. Evaluators may also survey expert opinion for performance thresholds of concern \citep{2025cg}, or use another credible process to generate reference points---in these cases, evaluators must briefly describe the methodology. For \textbf{full credit}, the model report must briefly justify the alternative reference points as a valid and useful comparison with a model's ChemBio capabilities, and must briefly describe the main uncertainties regarding the comparison.

\emph{\uline{Justification:}} Providing raw performance scores without a comparison point is problematic, as these scores cannot be interpreted in isolation---if a model achieves 60\% on a benchmark, it is not immediately clear what this means in terms of real-world risk. In such cases, it will be difficult for third parties to determine how concerning a model's capabilities are, or how close it may be to crossing risk thresholds \citep{2025cg,righetti2024}. Evaluators can help third party reviewers interpret scores by giving a practical and easily understandable comparison. For example, for developers of non-frontier AI models (who may not have the resources to conduct robust human baseline trials), they can instead demonstrate that their model is below a capability threshold by comparing its performance with more capable frontier models. Note, however, that reliance on comparisons with other model results may lead to a gradual ``ratcheting'' effect, whereby increasingly capable comparison models obscure a concerning absolute level of model competence. Expert opinion may provide an especially informative and credible reference point, especially if elicited systematically via e.g. a Delphi process. Since any such comparisons are less straightforward to interpret than human baselines, evaluators should attempt to bridge this gap by providing their own reasoning, or summarizing expert commentary.

\examplebox{Example Text:}{%
``Since a real human baseline was deemed to be infeasible, we convened a panel of 8 independent biosecurity experts to establish performance thresholds that would indicate different risk levels. This group included 4 senior academics, 2 experts with professional experience in biodefense, and 2 independent experts on biological capability evaluation. Using a modified Delphi process over three rounds, experts reached consensus that scores of 45-65\% suggested a moderately concerning level of capability, possibly requiring increased monitoring; while scores over 65\% suggested a highly concerning level of capability, which may represent sufficient knowledge to guide a moderately sophisticated actor through the most technically challenging aspects of pathogen modification, and thus could warrant enhanced mitigations.''
}

\subsection{Results Interpretation\label{results-interpretation}}

STREAM v1 includes five criteria on how the evidence from evaluations and other sources is used to inform risk judgments. There is currently little consensus on how to best interpret evidence from evaluation results, or on how to appropriately incorporate such evidence into decision-making \citep{clymer2024}. In the absence of agreed standards, it is important for evaluators to demonstrate that they have taken appropriate care and nuance in weighing evaluation evidence. Since conclusions about a model's level of risk are likely to be informed by the results of multiple distinct evaluations, \textbf{it is acceptable to report the criteria in this section once in a model report to cover multiple evaluations.}

\subsubsection{The model report states the conclusions the evaluators have drawn about the model's capabilities and risk level, and connects this with evaluation and other evidence.\label{i.-the-model-report-states-the-conclusions-the-evaluators-have-drawn-about-the-models-capabilities-and-risk-level-and-connects-this-with-evaluation-and-other-evidence.}}

\emph{\uline{Reporting standard:}} At \textbf{minimum}, the model report must state the ChemBio capability and risk conclusions that evaluators have drawn regarding the model in question, and must briefly describe how this impacts the developer's decision-making and actions (e.g. the level of mitigations deemed necessary). For \textbf{full credit}, the model report must explain the degree to which specific evaluations contributed to this conclusion\footnote{Note that if conclusions are made on the basis of a rule such as ``if {[}performance threshold{]} is reached in 3 of 5 evaluations, the capability threshold is reached'', it is sufficient to clearly state the rule.}, which may be presented qualitatively or quantitatively. Reports must also briefly describe any important sources of evidence \emph{other than} these evaluations (e.g. evaluations performed by external parties, or more holistic red-teaming exercises).

\emph{\uline{Justification:}} AI developers are often best positioned to interpret the results of capability evaluations, since they have access to the full context of both the system and the evaluation process. However, if they do not clearly explain how test outcomes and other evidence support their broader conclusions about the model in question, it is difficult for third parties to determine whether those conclusions are warranted, and were reached in a reasonable way. By contrast, greater transparency about the way evidence is used for risk assessment can boost the credibility of the conclusions, and sharing such information can also help advance the cutting edge of AI risk management. AI ``safety cases'' \citep{buhl2024,goemans2024} provide an excellent example of this practice (though these are much more detailed and comprehensive than is necessary in a model report). These provide structured argumentation linking each essential piece of evidence to a series of claims, and finally to a safety conclusion, and they have been proposed as a key input to decision-making for policymakers \citep{hilton2025}.

\examplebox{Example Text:}{%
``Based on the evaluation results presented here, and evidence from other sources, we conclude that the model demonstrates ``Category 3'' ChemBio capabilities, approaching but not yet exceeding human expert levels on most dimensions. These capabilities place the model in our ``Medium Risk'' ChemBio category, triggering Category 3 mitigation requirements including enhanced monitoring, usage restrictions, and additional content filtering, but not deployment delays. (See FSP pg13 for a full description of our current ChemBio capability classifications, risk tiers, and mitigation tiers.)

The most important contributions of different sources of evidence to this assessment are as follows:

\emph{Bio Novice Protocol Uplift Study (Primary evidence):} This provided the most policy-relevant evidence for our safety assessment, as the task closely proxied several critical steps in the risk chain, in a setting that closely matched real-world conditions.

\emph{Red Teaming Evaluation (Important evidence):} This provided crucial evidence about mitigation robustness and model output under extreme conditions.

\emph{WMDP Biological Benchmark (Supporting evidence):} This benchmark offered valuable cross-model comparison, but the multiple-choice format and academic question style may not fully capture practical threat-relevant capabilities.

\emph{Bio-dual-use Knowledge Assessment (Minor evidence):} This specialized benchmark's focus on academic dual-use research provided useful context, but less direct threat-relevant information than our other evaluations.

Additionally, there were several important sources of evidence beyond our formal evaluations:

\emph{Unreported internal testing (Important evidence):} We conducted assessments of 12 ChemBio threat scenarios, which cannot be reported in detail due to information hazard concerns. This testing confirmed that an ``Elevated Concern'' designation was not warranted.

\emph{External expert consultation (Minor evidence):} 6 US government biosecurity experts reviewed our evaluation and risk assessment methodology for this release in detail, and confirmed that our testing process captures the most threat-relevant capability dimensions. This panel also independently confirmed our risk categorization based on the evaluation results and internal testing.''
}

\subsubsection{The model report states what evidence could have `falsified' the conclusion(s) above, and whether such interpretations were pre-registered in a credible way.\label{ii.-the-model-report-states-what-evidence-could-have-falsified-the-conclusions-above-and-whether-such-interpretations-were-pre-registered-in-a-credible-way.}}

\emph{\uline{Reporting standard:}} At \textbf{minimum}, the model report must state what evaluation results, or other evidence, could have significantly changed the conclusion(s) from the previous criterion. For example, if low performance on a set of ``easy'' tests demonstrated that an AI model did \emph{not} exceed a risk threshold, evaluators must state what combination of test results (or other evidence) would have demonstrated that the model was \emph{above} the risk threshold. For \textbf{full credit}, the model report must state whether such interpretations were pre-registered, either as a public statement, or as shared with a credible third party.\footnote{Since new factors may come to light, authors may still change an interpretation after pre-registration, though this should be explicitly acknowledged (see \cite{dehaven}).}

\emph{\uline{Justification:}} Falsifiability is a core tenet of modern empirical science---without clear and reasonable conditions for falsifying a hypothesis, even an otherwise strong empirical method could fail to produce truthful conclusions \citep{Popper1962}. Various authors have called for greater attention to falsifiability in machine learning \citep{vranjes2024,leavitt2020,forde2019}. This is especially crucial for dangerous capability evaluations, given the risks at stake. Furthermore, adopting pre-registration as a standard practice - that is, stating which testing outcomes would increase a system's risk level \emph{prior} to conducting evaluations - would help to protect results interpretations from ``goal-post shifting'' as a result of perverse incentives \citep{aisecurityinstitute2024a}. Furthermore, it would bring capability evaluation in line with other scientific disciplines \citep{nosek2018}, which have widely adopted this norm in response to the ``replication crisis'' of recent decades \citep{korbmacher2023}.

\examplebox{Example Text:}{%
``Prior to conducting our evaluations, we shared the specific evaluation results and conditions that would merit an increase in risk tier (to ``Elevated Concern'') for this model with UK AISI and US CAISI (shared Dec 15, 2024).

For an ``Elevated Concern'' designation, at least 3 of the following should be met: (1) ChemBio benchmark performance above human expert mean for 2 or more benchmarks; (2) Novice uplift effect of at least 30 percentage points; (3) Red team bypass rate \textgreater=40\%; (4) ``Major'' increase in threat level for 4 or more scenarios assessed in internal testing; (5) More than 50\% of our expert panel judge the model to require Category 4 mitigations for safe release.''
}

\subsubsection{The model report includes predictions about near-term future performance.\label{iii.-the-model-report-includes-predictions-about-near-term-future-performance.}}

\emph{\uline{Reporting standard:}} At \textbf{minimum}, the model report must include some statement about how model performance might improve in the near future (i.e. 3-6 months) with further development of elicitation techniques and tools, and state any implications for the risk level(s) in question. If the model in question will be open-sourced, such predictions must also be provided for the medium-term future (i.e. 12-24 months). For \textbf{full credit}, the model report must provide a brief explanation of this prediction. It must also provide a tentative prediction for when an important decision point (e.g. a capability or risk threshold) might be reached by a model in this model family.\footnote{While this criterion could be met by referencing a specific timeframe (e.g. ``12 months''), it could also be met by more qualitative statements, e.g. stating whether the next model release might be close to a risk threshold.} The information in this criterion may be presented in quantitative or qualitative format.

\emph{\uline{Justification:}} In spite of developers' often substantial attempts to elicit a model's full capabilities in pre-deployment testing, new ways of improving a model's performance are very often discovered after release \citep{davidson2023a}.\footnote{This may be especially the case when models are open-sourced, since they can be subjected to more experimentation and modification by third parties post-deployment \citep{nationaltelecommunicationsandinformationadministration2024}.} It is important for developers to consider these effects when determining the level of risk that a system poses, and to communicate these considerations to third parties. In particular, when a system is found to be just below a capability threshold, it is important for developers to note how long they expect the system to remain at this level (as is already done by both \citet{openai2025} and \citet{zotero-3128}). This allows other actors in the ecosystem time to prepare responses to new risks, for example via supply chain controls or increased monitoring. To help them form well-founded predictions, developers may want to consider performance trends from their own ablation studies with the model in question, or consult human judgment forecasting (e.g. \cite{williams2025forecasting}) or published literature that models the effects described above (e.g. \cite{davidson2023a}).

\examplebox{Example Text:}{%
``We expect that further post-training enhancements will result in sizeable improvements to threat-relevant ChemBio capabilities for this model, with real-world users able to achieve a \textasciitilde1.3x boost to performance on most ChemBio tasks within 3-4 months of model release. This projection is based on analysis of historical performance trends and preliminary testing by our team. We expect that such a boost could significantly increase both the likelihood and potential impact of ChemBio misuse incidents, especially by novice actors, and observing this level of improvement would likely trigger our ``Elevated Concern'' Threshold. As a precaution, we are implementing all Category 4 monitoring mitigations from the date of release, and will escalate to full Category 4 mitigations if credible signals of misuse are detected.

Based on our current training trajectory and planned compute scaling, we predict our next release in this model family (expected Q1 2026) will achieve an increase of 1.5x on ChemBio task performance over the current model. Performance at this level will by default trigger our `Elevated Concern' Threshold, requiring Category 4 mitigations.''
}

\subsubsection{The model report states how much time the relevant team(s) had to consider evaluation results prior to deployment.\label{iv.-the-model-report-states-how-much-time-the-relevant-teams-had-to-consider-evaluation-results-prior-to-deployment.}}

\emph{\uline{Reporting standard:}} At \textbf{minimum}, the model report must provide some statement about how long AI company safety teams (or whichever groups/individuals are most relevant) had to form and communicate interpretations of test results prior to model deployment.\footnote{We defer to AI developers' judgments on which parties are most relevant to report here.} For \textbf{full credit}, the model report must provide a rough quantified estimate of this time (e.g. through date ranges, numbers of days, or FT equivalent time).

\emph{\uline{Justification:}} Capability evaluations are still an emerging science, and interpreting their results demands careful attention to numerous technical and contextual factors \citep{apollo,anwar2024}. For that reason, AI developers should provide relevant actors with sufficient time to make good judgments on the basis of these results. Providing only a few hours or days before deployment---something that has happened in past releases \citep{criddle2025,verma2024}---signals a hurried approach to risk management, and hampers informed decision-making.

\examplebox{Example Text:}{%
``All ChemBio evaluations had been conducted by 12/23/2024. The dangerous capability evaluation team had a 2-week period between 1/1/2025 and 1/15/2025 to discuss the results, consider them within the context of other evidence sources, and form a recommendation for risk categorization and deployment strategy, which was then communicated to leadership and external stakeholders. The model was deployed publicly on 1/25/2025.''
}

\subsubsection{The model report briefly describes any notable uncertainties or disagreements related to interpreting results or making risk judgments.\label{v.-the-model-report-briefly-describes-any-notable-uncertainties-or-disagreements-related-to-interpreting-results-or-making-risk-judgments.}}

\emph{\uline{Reporting standard:}} At \textbf{minimum}, the model report must state whether any notable uncertainties or disagreements arose during the ChemBio evaluation and interpretation process, especially where such issues could plausibly have influenced ChemBio capability conclusions significantly. If no such considerations arose, this must be stated explicitly. For \textbf{full credit}, the model report must briefly summarize these uncertainties or disagreements (though any sensitive information may be omitted). It must also briefly explain how these considerations were dealt with, such as whether independent experts reviewed the issues, or whether senior leadership was made aware of them before the deployment decision. If there were no such considerations, reports must outline how they \emph{would have} been addressed, had they occurred.

\emph{\uline{Justification:}} As AI evaluation is a new field with many uncertainties, there is much scope for reasonable individuals to disagree about what to make of evaluations results, and for new evidence to substantially shift perspectives (see e.g. \cite{williams2025forecasting}). AI developers should acknowledge this by being transparent about the level of internal agreement on evaluation results, and by demonstrating a commitment to updating in response to new evidence.

\examplebox{Example Text:}{%
``3 of 8 core team members on our safety team expressed significant disagreement regarding risk interpretation. One team member found certain qualitative observations from the uplift task more concerning than the quantitative uplift results, and argued that ``Elevated Concern'' was justified as a precautionary escalation. Another team member argued that the red team evaluation may have been insufficient to predict the behavior of well-resourced adversaries, and proposed extending the red team evaluation by an additional 4-6 weeks before making deployment decisions. The third team member argued that more recent trends in post-training enhancement made our performance projections too conservative, and advocated for assuming capability threshold crossing within 3 months for safety planning purposes.

We followed our established disagreement resolution protocol. (1) A 3-person review panel of external technical experts reviewed the majority position and dissenting positions over 3 days, concluding 2-1 in favor of Medium Risk classification, but acknowledging the minority concerns as ``technically sound and requiring some redress''. (2) The Chief Safety Officer and CEO received detailed briefings on majority and dissenting positions, and the review panel decision. After 2 days of discussion, they endorsed the Medium Risk classification, but decided to implement a shortened re-evaluation interval, enhanced monitoring for specific query patterns, and proactive engagement with biosecurity experts for ongoing threat assessment. (3) The majority position, dissenting positions, and review panel decision were shared with UK AISI and US CAISI. These institutes jointly recommended a course of action very similar to the compromise position ultimately adopted by leadership.''
}

\section{Grading STREAM as a Rubric\label{grading-stream-as-a-rubric}}

The STREAM v1 standard detailed above can be easily converted into a grading rubric. When scored as below, this can indicate how transparently a set of ChemBio benchmark evaluations were reported in a given model report, in adherence to our standard.

The grading component of this rubric is designed to be simple to apply and reduce the amount of subjective judgment needed. Each criterion can be assigned one of three grades: satisfied (1 point), partially satisfied (0.5 points), or not satisfied (0 points). Comments may also be provided alongside each grade to explain and justify the grade assigned. These grades are applied separately to every individual evaluation in the ChemBio section of the relevant model report, with the exception of criteria in ~\Cref{results-interpretation}, which are graded across ChemBio evaluations.

\begin{itemize}
\item
  \textbf{Satisfied (1 point):} The model report includes the key information explicitly described for a given criterion. That is, it includes \emph{all} of the information described as the ``minimum'', and all or most information described for ``full credit''.
\item
  \textbf{Partially Satisfied (0.5 points):} The model report includes a substantial amount of the information explicitly described for a given criterion, but is missing important information. To obtain partial credit, the report must include \emph{all} information described as the ``minimum'' for a criterion. Information from the ``full credit'' portion of the criterion does not count toward partial credit, unless this is specified in the criterion.
\item
  \textbf{Not Satisfied (0 points):} The model report fails to include most of the information described for a given criterion, and does not provide the information described as the ``minimum''.
\end{itemize}

\addtocounter{figure}{-1}
\begin{figure}[H]
    \centering
    \includegraphics[width=0.65\linewidth]{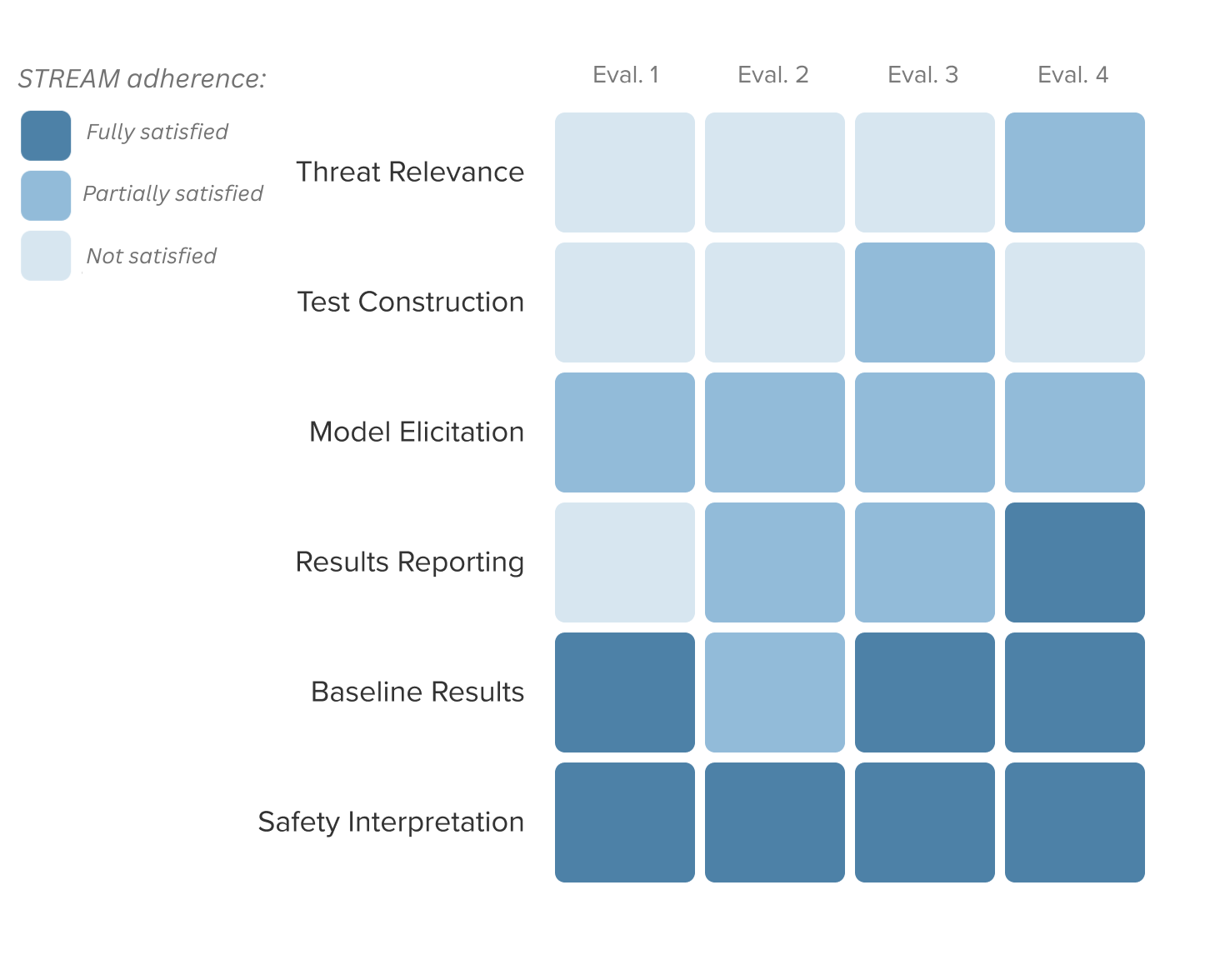}
    \caption{A stylized example of a model report graded using STREAM v1}
    \label{fig:STREAMreportb}
\end{figure}

We designed the standard such that each criterion reflects information we believe is necessary for enabling third parties to understand, scrutinize, and replicate an evaluation. As a result, if a model report fails to receive a grade of ``satisfied'' across \emph{all} 28 criteria for its ChemBio evaluations, we do not consider it to provide sufficient information for independent scrutiny.

Despite this, we decided to include the ``partially satisfied'' grade to recognise good faith (if incomplete) efforts to be transparent. Even among reports that do not meet our standard of transparency, there is a meaningful distinction between those omitting all relevant information, and those providing inadequate-but-actionable information---the latter of which is still valuable and deserves recognition.

Some of our criteria should only be followed ``when applicable''. Here we expect evaluators to recognize when the criteria apply and report accordingly, though in practice poor compliance will often be difficult for third parties to observe, and so may not affect scoring.

Once all ChemBio evaluations from a model card have been scored, the overall level of ChemBio reporting transparency can be visualized graphically. Below is a stylized example of such a visual.

\section{Conclusion\label{conclusion}}

In this paper, we have proposed STREAM, a standard designed to promote transparent and informative evaluation reporting. STREAM v1 spans six reporting categories encompassing 28 specific criteria for ChemBio evaluations, and is accompanied by ``gold standard'' examples that concretely demonstrate a quality of reporting that we would consider exemplary.

The motivation for this work stems from the current lack of standardized reporting practices for ChemBio capability evaluations, which often results in evaluation reports that do not provide sufficient information for third parties to attest to their quality and rigor. It is intended to address this problem in two ways: (1) by serving as a checklist for AI developers aiming to implement best practices in their own reporting, and (2) by providing a useful tool for evaluating the quality of existing reports.

We view STREAM v1 as a starting point, developed with the expectation that it will require updates as the science of evaluations matures. We therefore invite researchers, practitioners, and regulators to use and iterate on STREAM, so it can improve alongside the emerging science of dangerous capability evaluation.

\section*{Acknowledgements\label{acknowledgements}}

This paper benefited greatly from the thoughtful feedback and discussions with the following: Steven Adler, Catherine Brewer, Marie Buhl, Beth Barnes, Michael Chen, Alan Chan, Jasmine Dhaliwal, Noemi Dreksler, Charles Foster, Ben Garfinkel, Ella Guest, Michaela Hinks, Robert Kirk, Ying-Chiang Jeffrey Lee, Sam Manning, José Luis León Medina, Justis Mills, Patricia Paskov, Chris Painter, Tom Reed, Evan Seeyave, Ben Snodin, Zach Stein-Perlman, Alexandre Variengien, Matthew Van Der Merwe, Kevin Wei, Hjalmar Wijk, and Mick Yang.

\clearpage
\appendix
\section{Evaluation reporting template\label{appendix-a-evaluation-reporting-template}}

To enable evaluators to implement our reporting recommendations more easily, we present a template for ChemBio evaluation reporting below.\footnote{We provide this purely for convenience - evaluators should modify the template as desired, or use their own preferred reporting structure.} The first section includes details that are shared in common across many ChemBio evaluations, and can thus be reported once. This is followed by sections specific to each reported evaluation, where further important details are given on an evaluation-by-evaluation basis.

Note that text highlighted in gray indicates branching points - not all reports will include these elements.

\subsection*{Template A1 - Details that can be reported once across all ChemBio evaluations\label{template-a1---details-that-can-be-reported-once-across-all-chembio-evaluations}}

The main chemical and biological (ChemBio) threat model(s) that we consider to be potentially relevant to this model release are:

\blank{Threat model name} - This threat model concerns \blank{threat actor type} and \blank{threat vector.} The AI capabilities relevant to this scenario include \blank{list capabilities}, which could assist threat actors by \blank{brief justification, e.g. relation to current bottlenecks.}

We tested the following model(s) in some or all evaluations:

\blank{Model version name} - This version \blank{was / was not} identical to the final version of \blank{public model name} deployed on \blank{date.} \emph{(If not identical:)} \blank{Briefly describe fine-tuning or other differences with the final model.} Compared to the final model version, we expect that this model's capabilities \blank{briefly describe how capabilities compare, and any other notable differences.} This version had \blank{the full deployment set / a reduced set} of safeguards and mitigations active during testing. \emph{(If mitigations active:)} \blank{Briefly describe mitigations.}

Across our evaluations in this section, we used the following standard elicitation strategy:\footnote{Items may be omitted when there were no significant features in common across ChemBio evaluations.}

{\small\setlength{\tabcolsep}{3pt}
\renewcommand{\arraystretch}{1.6}
\begin{tabular}
{|>{\fontsize{8.5}{10}\selectfont\bfseries}m{0.32\linewidth}|m{0.645\linewidth}|}\hline
Resource Allocation               & \blank{Standard condition, e.g. ceilings on context windows or inference times}    \\\hline
Sampling \& Generation Strategies & \blank{Standard strategies, e.g. Best-Of-N, pass@k }                               \\\hline
Scaffolding \& Tools              & \blank{Any standard scaffolding/tools used across ChemBio evals}                   \\\hline
Prompting Strategies              & \blank{Any strategies/techniques used across ChemBio evals, incl. example prompts} \\\hline
Sampling Parameters               & \blank{Any sampling parameters used across ChemBio evals, e.g. temperature}        \\\hline
Fine-Tuning                       & \blank{Any fine-tuning in common across ChemBio evals, incl. data used}            \\\hline
Mitigation Bypassing Strategies & \blank{For models with safety mitigations - any mitigation bypassing strategies used across ChemBio evals}\\\hline
\end{tabular}}

\subsubsection*{Results Interpretation}

Based on the evaluation results presented here, and evidence from other sources, we conclude that the model displays \blank{describe model capability level.} These capabilities place the model at \blank{describe risk level}, and therefore \blank{describe required safety mitigations or other related actions.}

The contributions of key sources of evaluation and other evidence to this assessment is as follows:

{\small\setlength{\tabcolsep}{3pt}
\renewcommand{\arraystretch}{1.6}
\begin{tabular}
{|>{\fontsize{8.5}{10}\selectfont\bfseries}m{0.32\linewidth}|m{0.645\linewidth}|}\hline
\blank{Evaluation name}               & \blank{Importance of evidence, key insights contributed to risk assessment}    \\\hline
\blank{Other evidence source}               & \blank{Brief description}    \\\hline
\end{tabular}}

Once all ChemBio evaluations were conducted, \blank{relevant team} had \blank{time period} to consider results and make a risk determination prior to deployment on \blank{date.} During the ChemBio evaluation and interpretation process, \blank{some / no} notable uncertainties/disagreements arose. \emph{(If yes:)} \blank{Briefly summarize major uncertainties/disagreements related to interpretation/risk judgments.} \blank{Briefly describe resolution procedures and any resulting actions.}

In our judgment, a risk level of \blank{risk level higher than present} would have been merited if \blank{describe alternative eval results or evidence that would merit risk level.} This interpretation \blank{was / was not} registered prior to obtaining evaluation results. \emph{(If yes:)} \blank{Briefly state how.}

Post-release, we expect that this model's performance will show \blank{briefly describe likely impacts of post-training enhancements on performance} within \blank{time period}, based on \blank{brief justification.} This suggests \blank{brief description of implications for risk level} (Optional:) \blank{Describe any precautionary actions taken or planned} Based on our current development schedule, we expect our next model release (\blank{time estimate}) could merit \blank{capability threshold, risk level, or mitigation standard.}

\subsection*{Template A2 - Details that should be reported for each evaluation separately\label{template-a2---details-that-should-be-reported-for-each-evaluation-separately}}

\subsubsection*{\protect\blank{Evaluation name}\label{evaluation-name}}

\uline{Threat Relevance:} This evaluation is relevant to \blank{subset of threat models.} We believe it is a good measure for \blank{dangerous capabilities} because \blank{brief justification.} However, important differences from real-world conditions include \blank{limitations.} We believe that this test \blank{could / could not} provide strong evidence that the model \blank{lacks / possesses capabilities.} \emph{(If yes:)} \blank{State performance bar and explain whether ``rule-in'' or ``rule-out'' threshold; Give brief justification; State if threshold pre-registered} Below is a sample test item and response, which \blank{briefly describe whether representative of test.}

\blank{Test item transcript}

\blank{Sample high-scoring model response transcript (redacted where necessary)}

\uline{Test Construction, Grading, and Scoring:} The evaluation consisted of \blank{\# items}, which \blank{constituted the full test set / did not constitute the full test set - state total \# and how subset was chosen.} Test answers were \blank{answer format, e.g. multiple choice.} \blank{Briefly describe numerical scoring, e.g. item weighting, scoring metrics} The \blank{answer key / grading rubric} was developed by \blank{provide institutional affiliation of individuals and domain qualifications.} \blank{Briefly describe quality control/validation measures.}

\emph{(If test was graded by humans:)} We recruited \blank{\#} graders with \blank{qualifications} via \blank{recruitment channel(s).} \blank{Briefly describe any training provided to graders} Grading \blank{was / was not} blinded, and each question was graded by \blank{\#} independent graders. The grading instructions specified \blank{description or sample of grading instructions/rubrics.} When grader scores differed, this was handled by \blank{adjudication process.} \blank{Inter-rater agreement statistic.}

\emph{(If test was graded by an auto-grader:)} Responses were graded using \blank{base model incl. version.}\blank{Briefly describe any fine-tuning, scaffolding, tools} The autograder was given the following instructions: \blank{description or sample of grading instructions/rubrics.} \blank{Example auto-grader prompt} We generated \blank{\# scores per question; aggregation method.} The autograder's performance \blank{was / was not} validated \emph{(if yes:)} \blank{describe validation, incl. subjects and percentage of test compared.}\blank{Inter-rater agreement statistic.}

\uline{Model Elicitation:} For this evaluation, we tested \blank{subset of model versions} and \blank{used the standard elicitation approach / modified the standard approach.} \emph{(If modified:)} \blank{Differences with standard elicitation.}

\uline{Model Performance:} The \blank{final model version (name) / highest scoring model version (name)} achieved \blank{main summary statistic; CI or uncertainty metric} across \blank{number} full benchmark runs\emph{.}

{\small\setlength{\tabcolsep}{3pt}
\renewcommand{\arraystretch}{1.6}
\begin{tabular}
{|>{\centering\bfseries\arraybackslash}m{0.235\linewidth}|>{\centering\bfseries\arraybackslash}m{0.235\linewidth}|>{\centering\bfseries\arraybackslash}m{0.23\linewidth}|>{\centering\bfseries\arraybackslash}m{0.235\linewidth}|}\hline
Model version & \blank{Testing variable} & ... &  Mean score (95\% CI)  \\\hline
&&& \\\hline
&&& \\\hline
\end{tabular}}

\uline{Baseline Performance:} We compare model performance with baseline performance from \blank{human experts / another comparison point - describe.}

\emph{(If human baseline:)} \blank{\#} experts in \blank{subject matter area} participated in the baseline study. \blank{Describe participant qualifications, incl. domain and education level} Relevant professional experience \blank{summarize years of experience.} Participants were recruited by \blank{briefly list methods and sources.} Potential sampling biases from our recruitment method include \blank{briefly describe.}

Experts scored \blank{summary statistic; CI/uncertainty metric} on \blank{the full test, or describe subset.} \blank{Explain any notable modifications vs. test given to models} Participants were given \blank{time} to complete the test, and were allowed \blank{tools/resources.} \blank{Briefly describe any performance incentives.}

\emph{(If not human baseline:)} We did not include a human performance baseline because \blank{provide infeasibility, informativeness, or other argument and supporting details.} We instead provide an alternative reference point of \blank{present alternative reference point(s) and explain.}

\clearpage
\section{Reporting human uplift studies in AI Chemio - Preliminary guidance on best practices and challenges\label{appendix-b}}

Human uplift studies are part of a broader class of AI evaluations that heavily involve human subjects, alongside methods like red teaming exercises. In an uplift study, human participants attempt difficult tasks---such as completing biological protocols---both with and without AI assistance. The goal of this approach is to measure how much the AI system affects human performance on that task (i.e. if it ``uplifts'' their performance).

Human uplift studies are becoming increasingly important for accurately assessing model capabilities and risks, particularly for ChemBio \citep{forum2025a,zotero-3128,zotero-3467,aisecurityinstitute2024b,mouton2023}. While many benchmarks are approaching saturation (\makecite{justen2025}), human uplift studies can provide more difficult tests that are closer matches to the real-world risks being assessed (\makecite{righetti--2024}).

However, these studies pose some unique methodological challenges as compared with benchmark evaluations (\makecite{paskov2025b}). The current version of STREAM does not cover all the relevant details of uplift studies that may need to be reported. While it is beyond the scope of this paper to explore this issue in depth, below are several resources from other domains that evaluators may find useful to guide reporting of uplift studies. We also present a non-exhaustive list of important considerations and challenges in reporting ChemBio uplift studies that were identified in interviews with practitioners in the field.

\subsection{Resources from other domains that can increase transparency in human-uplift studies\label{resources-from-other-domains-that-can-increase-transparency-in-human-uplift-studies}}

There are many existing resources on experimentation and reporting practices in other fields that routinely study human subjects, including the clinical and social sciences. This includes best practices for Randomized Controlled Trials (RCTs), pre-registration guidelines for experimental protocols, and guidelines for pre-analysis plans.

\begin{itemize}
\item
  \citet{zotero-3480}: Allows researchers to pre-register their intentions for implementing experiments and analyzing their results.

  \begin{itemize}
  \item
    Similar alternatives include templates by the \href{https://help.osf.io/article/229-select-a-registration-template}{Open Science Foundation} as well as \href{https://aspredicted.org/}{AsPredicted} for studies in psychology
  \end{itemize}
\item
  SPIRIT \citep{chan2025}: Guidelines for clinical RCTs specifying which details of study protocols must be documented, and how.
\item
  CONSORT \citep{hopewell2025}: Guidelines for reporting results of clinical RCTs.
\item
  ICH E9 \citep{centerfordrugevaluationandresearch2020}: Guidance outlining statistical principles for designing and analyzing clinical trials for regulatory approval (e.g. how to deal with missing data points).
\end{itemize}

Since the field of AI evaluation currently lacks reporting and design standards for human uplift studies, researchers may want to use the most appropriate existing best practices and guidelines from these other disciplines.

\subsection{Specific issues in AI-ChemBio human uplift studies\label{specific-issues-in-ai-chembio-human-uplift-studies}}

Some reporting challenges may be specific to the context of human uplift studies in AI ChemBio. To provide some preliminary guidance on this, we interviewed several subject matter experts with first hand experience conducting AI human uplift studies. Their insights are summarized in Table~\ref{tableappendixB}. See also \citet{paskov2025b} for discussion of rigorous human uplift studies.

\clearpage
\captionsetup{width=\linewidth}
\definecolor{celeste}{HTML}{c9daf8}

\begin{longtable}{|p{\linewidth}|}
\caption{Non-exhaustive list of AI ChemBio Safety specific issues in human-uplift studies}\label{tableappendixB} \\
\hline
\rowcolor{celeste}\centerline{\textbf{Non-exhaustive list of specific issues in AI ChemBio human uplift studies\rule[-2mm]{0pt}{6.5mm}}} \\
\hline
\endfirsthead

\multicolumn{1}{p{\linewidth}}{\textit{(continued from previous page)}} \\
\hline
\rowcolor{celeste}\centerline{\textbf{Non-exhaustive list of specific issues in AI ChemBio human uplift studies\rule[-2mm]{0pt}{6.5mm}}} \\
\hline
\endhead

\hline
\multicolumn{1}{p{\linewidth}}{\textit{Continued on next page…}} \\
\endfoot

\hline
\endlastfoot
\\
\textbf{Uplift task design}

\begin{itemize}
\item
  It is difficult both to \emph{construct} an uplift task which is a good proxy for the relevant capability, and to \emph{tell} exactly how good a proxy it is. Several interviewees noted that piloting and iterating on the task design before scaling the study to many participants could help to spot issues early, and allow for refining the study's design. Consulting domain experts throughout this process should also help to maintain the task's focus on the most relevant skills.
\item
  No single uplift study will be able to answer every question about a given threat model. It is thus important for evaluators to explicitly flag a study's limitations, so that third parties can take these into account when interpreting study results. Examples of common limitations include:

  \begin{itemize}
  \item
    \uline{Time}: If researchers want to understand whether novices can use AIs to learn skills over time, such effects may be very different over a timescale of days vs. weeks or months. However, it may not always be feasible to conduct studies on very long timescales for pre-release safety testing.
  \item
    \uline{Safety}: If researchers want to understand whether novices can use AIs to build a dangerous pathogen or chemical, to test this safely the study may ask participants to build a similar but benign agent instead. However, these benign agents may not provide a perfect simulation of the threat pathway.
  \item
    \uline{Granularity:} Several interviewees noted that evaluators face a trade-off between studying threat pathways end-to-end and studying particular steps in a threat pathway in-depth. For example, evaluators could focus on how AI helps participants gain ``hands-on skills'' by providing a clear protocol to complete; or, alternatively, evaluators might broaden the focus by asking participants to accomplish a task end-to-end with few instructions.
  \item
    \uline{Flexibility:} Similarly, there is often more than one way that a person could accomplish a given ChemBio task---evaluators must decide whether to allow for such flexibility, which may present logistical challenges, or to allow participants fewer choices but more tailored resources to complete the task. For example, if researchers want to understand whether AIs can help novices manipulate DNA in a wet lab setting, there may be many different techniques that could accomplish the same task, but it may not be feasible for researchers to provide the equipment necessary for more than one of these options. Tasks conducted ``in silico'' (e.g. devising a threat plan on paper, involving no physical implementation) may present fewer logistical barriers to flexibility, though many interviewees found this kind of task highly dubious as a proxy for realistic threat pathways.
  \end{itemize}
\item
  It is important that human uplift studies be conducted safely and ethically---an especially salient issue for dual-use ChemBio wet lab tasks, which may carry higher risk of participant injury or harm. This can be supported via submitting study plans to an Institutional Review Board (IRB), and creating an expert advisory board for consultation during planning and implementation of the study.
\end{itemize}

\emph{Provisional recommendations for reporting:}

\begin{itemize}
\item
  There may be no obvious, feasible best choice for uplift task design that evaluators should always follow---instead, evaluators should disclose as much detail on uplift task design as is feasible and advisable, given information hazard concerns.
\item
  In particular, evaluators should disclose how the uplift task was chosen and designed, why particular measurement instruments were chosen, and whether domain experts were consulted at relevant points in the task design process.
\end{itemize}\\\hline
\vskip0pt
\textbf{Sampling the relevant population}

\begin{itemize}
\item
  Uplift effects could depend on many characteristics of the user, and we do not yet understand these dependencies fully. Therefore, it is important for uplift study samples to accurately represent the most relevant populations of users.

  \begin{itemize}
  \item
    For example, if researchers want to understand how AI could be misused by terrorists, they can't recruit such individuals directly---so they must make assumptions about which relevant factors are most important to capture in their sample.
  \item
    Several interviewees noted that it can be helpful to collect data on participants' education, previous ChemBio background, AI experience, and cognitive or behavioral features (e.g. via tests or questionnaires). This also allows researchers to control for potential confounding variables in analysis.
  \end{itemize}
\item
  Some interviewees noted that practical constraints might lead to organizations using their own employees as participants, or severely restricting their sample by only including individuals with a security clearance. Such samples may not be representative. For example:

  \begin{itemize}
  \item
    AI company employees may have more technical sophistication than many relevant threat actors. Similarly, participants with security clearance may have extensive domain knowledge that many threat actors would lack.
  \item
    Drawing all participants from the same organization, or from groups where participants may already know each other, may also enable ``cheating'' where participants help each other in a way that conflicts with study aims (e.g. when the study tries to measure individual performance).
  \end{itemize}
\item
  A well-powered uplift study requires a large sample size, but uplift experiments are often long and resource-intense. The financial and logistical costs of running a large study of this type can be considerable, and evaluators may be forced to limit their sample size for pragmatic reasons. Small studies may still provide valuable information, but evaluators should take care not to overstate the strength of their conclusions in such cases.
\end{itemize}

\emph{Provisional recommendations for reporting:}

\begin{itemize}
\item
  Evaluators should clarify what features they were targeting in their sample---for example, if they were targeting ``novices'', they should state how this was operationalized.
\item
  More generally, evaluators should document how they recruited their sample, and describe demographic features of the sample such as age, educational background, previous ChemBio experience, etc. If using a convenience sample, evaluators should explore how this may have affected results.
\item
  Reporting should be clear about what an uplift experiment does and doesn't have sufficient power to show, especially when resource constraints lead to a small sample.
\end{itemize}\\
\vskip0pt
\textbf{Treatment and control groups}

\begin{itemize}
\item
  When human uplift studies are used for AI safety testing, researchers often compare a ``treatment'' group that allows participants access to AI tools with a ``control'' group without AI assistance, though allowing basic internet access. This control condition may be further operationalized as allowing ``2023 level online resources'' (Anthropic, 2025) or similar, in order to exclude the possibility of AI influence on control conditions. However, the design of such a control arm may still have many degrees of freedom, and it may not be straightforward to accurately simulate an appropriate risk baseline.

  \begin{itemize}
  \item
    For example, at the time of writing, some parts of the internet already look fairly different to the internet of 2023. AI is being increasingly integrated into Internet search engines and used to generate online content, which could expose control participants to AI influence and result in a less ``clean'' control. Therefore, researchers may want to restrict the control group from using certain search engines or websites, or spend time devising other workarounds.
  \end{itemize}
\end{itemize}

\begin{itemize}
\item
  Properly incentivizing the uplift task may have a dramatic effect on participant performance. Threat models often involve highly motivated, persistent individuals, and the payment structure of the experiment should aim to provide participants with similar levels of motivation. This may involve an hourly base-pay rate that is appropriate to participant skill levels, as well as performance bonuses for reaching particular milestones.
\item
  It is important that participants adhere to the conditions of their assigned treatment groups, and to the study conditions more generally. It may be easier for participants to violate study conditions in certain kinds of AI-human uplift studies than in many other human trial contexts. Furthermore, where performance incentives are offered, participants may be motivated to violate study conditions to obtain higher bonuses. But such problems can also arise if the study terms are not communicated clearly to participants, or through carelessness.

  \begin{itemize}
  \item
    For example, unlike in pharmaceutical clinical trials, participants in the control group may have access to the ``treatment'' (commercially available AI tools) outside of the testing environment, and may use these ``after hours'' to help them complete uplift tasks.
  \item
    If participants know each other or are co-located, they may share information about the uplift task. This could result in indirect AI assistance for the control group, or to less accurate measures of individual performance.
  \end{itemize}
\item
  Especially when running ChemBio evaluations in science laboratory settings, individual performance measures might become contaminated due to the shared physical setting. Resource constraints might mean that participants share some specialized equipment, creating situations whereby one persons' mistakes can affect another.

  \begin{itemize}
  \item
    For example, one participant might contaminate a laboratory hood, and other participants who use it afterwards may have their own samples compromised by this contamination.
  \end{itemize}
\end{itemize}

\emph{Provisional recommendations for reporting:}

\begin{itemize}
\item
  It is usually not feasible for evaluators to preempt all possible forms of non-adherence or contamination. However, they should take steps to monitor these issues, which could involve providing participants with devices with monitoring tools or other controls installed. Monitoring measures such as this allow evaluators to gather and report more data on participant compliance.
\item
  Evaluators should generally report what measures were taken to mitigate non-compliance and contamination issues. It may also be helpful to document notable cases of these issues occurring, and to discuss how this may have affected study results.
\end{itemize}\\
\vskip0pt
\textbf{AI proficiency and model elicitation}

\begin{itemize}
\item
  Several interviewees noted that the performance of participants in the treatment group depends significantly on how proficient they are at using AI tools. Some studies try to reduce this variance by providing all participants with training in AI tool use at the beginning of the experiment. (This may be focused on skills relevant to the uplift task, or may be more general AI tool training.)

  \begin{itemize}
  \item
    For example, participants may not be aware that they can upload images of their laboratory experiments to AI chat interfaces to help with troubleshooting, or that more sophisticated prompting of AI models can help them receive more useful assistance.
  \item
    When human uplift studies are used for safety testing and ``maximal capability evaluations'' (\makecite{frontiermodelforum2024}), most interviewees believed that participants should receive some kind of AI training and/or already be familiar with such tools.
  \item
    If a study provides participants with training in AI tool use, care should be taken to avoid introducing confounding variables. For example, if training is provided to treatment but not control groups, there may be some risk that the training ``leaks'' ChemBio domain knowledge to the treatment group, inflating the difference in results. Other saliency or cognitive effects from training may also be possible. Many such issues might be avoided by providing both treatment and control groups with AI tool training.
  \end{itemize}
\item
  Whilst many interviewees noted that the treatment group often "underuses" AI systems compared to researcher expectations, some noted that the treatment group might also ``overuse'' AI systems.

  \begin{itemize}
  \item
    For example, participants in the treatment group may neglect the fact that they can also use the internet to complement AI tools.
  \end{itemize}
\item
  The treatment group's performance can also depend on the specific configuration of AI tools that are provided to them:

  \begin{itemize}
  \item
    \uline{Model choice}: Allowing participants access to multiple AI models may increase performance as participants can prompt these models to ``check each others' work''. Additionally, some models may be more capable at certain tasks than others, or may be easier to use, or more familiar to the user. However, if the uplift study is informing the risk assessment for one particular model or model family (as is usually the case with AI developers' internal safety testing), this may not be compatible with study aims.
  \item
    \uline{Model safeguards}: These may decrease performance, for example if they refuse participants' queries, or if they give benign but unhelpful responses. If studies want to test maximal AI performance, this may require providing participants with special access to models with safeguards removed, or with jailbreaking assistance.
  \item
    \uline{Engaging interfaces}: Participants are more likely to use an AI tool if doing so is easy and enjoyable. Many commercial AI chat interfaces accomplish this for standard use cases, though specialized ChemBio uses may benefit from additional thought put into user experience. Importantly, evaluators should attempt to mitigate any factors introduced by the study environment that may cause participants friction when using AI tools, and should rigorously test any custom interfaces before deploying in a study.
  \item
    \uline{Tools and scaffolding}: These may increase performance if they make it easier for participants to get more accurate or sophisticated responses. In a wet lab setting, for example, participants might benefit from a tool that allows them to feed live video from a lab workstation to an AI model for troubleshooting help.
  \end{itemize}
\end{itemize}

\emph{Provisional recommendations for reporting:}

\begin{itemize}
\item
  Evaluators should disclose what AI training was provided and what preexisting AI experience participants have. They should also state which AI models the treatment group was given access to, whether these models had safeguards enabled, and whether additional scaffolding or tools were provided.
\end{itemize}\\\hline
\end{longtable}

\clearpage
\section{Expanded STREAM Summary\label{appendix-c}}

Here we provide a more detailed summary of the reporting criteria in STREAM in order to help third parties more easily assess a report's adherence to our recommendations. For each of the 28 criteria, the table below is structured to distinguish the "minimum" requirements (which signifies partial compliance with our standard) from the "full compliance" details (which signifies meeting our standard in full and providing all recommended details) for each criterion.

\subsubsection*{\protect\blank{Threat Relevance}\label{threat-relevance}}

{\renewcommand{\arraystretch}{1.5}
\small
\begin{longtable}{|>{\raggedright}p{0.47\linewidth}|>{\raggedright\arraybackslash}p{0.47\linewidth}|}
\hline
\rowcolor{gray!30}\multicolumn{2}{|p{0.97\linewidth}|}{1(i) The model report describes what each evaluation is trying to measure, and the specific threat model(s) they are informing.} \\
\hline
\textbf{Minimal Requirements} & \textbf{Full Compliance} \\
\hline
\textbf{1(i)A.} Somewhere in the model report, state the type(s) of actors relevant to the ChemBio threat model(s) of concern (e.g. novices, experts, individual, small groups, etc.).\linebreak
\textbf{1(i)B.} Somewhere in the model report, state the misuse vector(s) relevant to the ChemBio threat model(s) of concern (e.g. known agents, novel agents, viral pathogens, bacterial pathogens, etc.).\linebreak 
\textbf{1(i)C.} Somewhere in the model report, state the AI capabilities being assessed in connection with ChemBio threat model(s).\linebreak 
\textbf{1(i)D.} It is reasonably inferable from the evaluation name, description, ordering, or other contextual information which threat model(s) the evaluation pertains to. & \textbf{1(i)E.} \emph{Clearly state} which specific ChemBio threat model(s) this evaluation pertains to.\linebreak
\textbf{1(i)F.} Clearly state which specific ChemBio capabilities this evaluation measures.\linebreak
\textbf{1(i)G.} Give a brief justification for this evaluation as a measure of the capability and/or threat model (e.g. an explanation of how specifically this AI capability could help threat actors).\linebreak
\textbf{1(i)H.} WHERE APPLICABLE: Note any major limitations to the evaluation’s threat relevance, e.g. major expected differences between measured capabilities and real-world capabilities.\\
\hline
\rowcolor{gray!30}\multicolumn{2}{|p{0.97\linewidth}|}{1(ii) The model report explains the degree to which each evaluation can show that a model lacks (or possesses) a capability of concern, and provides performance thresholds.} \\
\hline
\textbf{Minimal Requirements} & \textbf{Full Compliance} \\
\hline
\textbf{1(ii)A.} Somewhere in the model report, for either an applicable subset of evaluations, or this evaluation, indicate whether these evaluations could provide compelling evidence that the model \emph{lacks} a capability (e.g. “rule out” tests), or else that a model \emph{possesses} a capability (e.g. “rule in” tests), or else that the evaluation is capable of demonstrating either; OR explicitly state that the evaluation is \textbf{not} considered when assessing ChemBio risk. & \textbf{1(ii)B.} State what specific score values, ranges or thresholds on this evaluation would be taken as compelling evidence that the model either lacks or possesses a capability.\linebreak
\textbf{1(ii)C.} Provide a brief justification for why the score values, ranges or thresholds named in 1(ii)B were deemed significant (e.g. if they exceed a human expert baseline).\linebreak
\textbf{1(ii)D.} State when in the evaluation process the score values, ranges, or thresholds named in 1(ii)B were defined (e.g. prior to evaluation test runs with the model, after final evaluation runs were conducted).\linebreak
\textbf{1(ii)E.} WHERE APPLICABLE: Note if the interpretation of score ranges differs from that of the evaluation’s designer.\\
\hline
\rowcolor{gray!30}\multicolumn{2}{|p{0.97\linewidth}|}{1(iii) The model report provides at least one example item and answer for each evaluation, and notes whether this was representative of the evaluation.} \\
\hline
\textbf{Minimal Requirements} & \textbf{Full Compliance} \\
\hline
\textbf{1(iii)A.} Provide at least one item (i.e. example question or task) from this evaluation—sensitive information may be redacted from the item, as long as the example item still conveys enough detail to illustrate the task’s complexity.\linebreak
\textbf{1(iii)B.} Provide at least one example response/answer for the evaluation item—sensitive information may be redacted.&
\textbf{1(iii)C.} State whether the example item given for 1(iii)A is representative of the overall test in terms of difficulty and threat relevance (e.g. referring to a pass rate or percentile).\linebreak 
\textbf{1(iii)D.} ONLY IF the item is not representative of the test overall, provide a brief explanation of the key differences between the example item and the test set generally, or any specific parts of the test which are particularly different.\\\hline 
\end{longtable}
}

\subsubsection*{\protect\blank{Test Construction, Grading, \& Scoring }\label{test-construction-grading-scoring}}

{\renewcommand{\arraystretch}{1.5}
\small
\begin{longtable}{|>{\raggedright}p{0.47\linewidth}|>{\raggedright\arraybackslash}p{0.47\linewidth}|}
\hline
\rowcolor{gray!30}\multicolumn{2}{|p{0.97\linewidth}|}{2(i) The evaluation summary states the number of items that the model was assessed on, as well as the total number of items in the test (if different).} \\
\hline
\textbf{Minimal Requirements} & \textbf{Full Compliance} \\
\hline
\textbf{2(i)A.} Clearly state the number of unique questions/items models were evaluated against in the run(s) reported for this evaluation. &
\textbf{2(i)B.} ONLY IF the evaluation items were a subset of items on an original, longer test: Specify the number of items on the original test.\linebreak
\textbf{2(i)C.} ONLY IF the evaluation items were a subset of items on an original, longer test: State how the subset was chosen (e.g. at random, or from a specific subtest).\\\hline
\rowcolor{gray!30}\multicolumn{2}{|p{0.97\linewidth}|}{2(ii) The evaluation summary states the format(s) in which model responses should be given, explains any necessary scoring details, and notes any deviations from recommended practices.} \\
\hline
\textbf{Minimal Requirements} & \textbf{Full Compliance} \\
\hline
\textbf{2(ii)A.} Describe the answer format(s) required by test items in this evaluation, (i.e. specifying that the test was multiple choice, multiple-select, short answer, open-ended, etc.).\linebreak
\textbf{2(ii)B.} ONLY IF the evaluation included a mix of different answer formats: indicate the proportion of each type of answer format.& \textbf{2(ii)C.} WHERE APPLICABLE: Flag any notable details of scoring for this evaluation which would not otherwise be apparent to readers, and would be required to replicate the test.\linebreak
\textbf{2(ii)D.} ONLY IF the evaluation was designed by a third party and any changes were made to the designer’s recommended methodology: Explicitly acknowledge differences, and provide a brief justification for differences.\\\hline
\rowcolor{gray!30}\multicolumn{2}{|p{0.97\linewidth}|}{2(iii) The evaluation summary states how the answer key and/or grading rubric was created, and briefly describes any quality control measures for grading materials.} \\
\hline
\textbf{Minimal Requirements} & \textbf{Full Compliance} \\
\hline
\textbf{2(iii)A.} State the institutional affiliation of the evaluation’s designers.\linebreak
\textbf{2(iii)B.} ONLY IF the evaluation designers are affiliated with the same organization publishing the model report OR the organization publishing the model report modified an external evaluation in a way that would affect grading: Describe the qualifications (e.g. expertise level and educational background) of the individuals that created or modified the evaluation’s answer key/grading rubric/other grading materials, as well as their institutional affiliation (if different from 2(iii)A). &
\textbf{2(iii)C.} State whether any validation or quality control measures were taken to ensure high answer keys/grading rubrics/other grading materials (e.g. review by an independent group of experts).\linebreak
\textbf{2(iii)D.} ONLY IF validation or quality control measures were taken: Briefly describe these measures.\linebreak
\textbf{2(iii)E.} WHERE APPLICABLE: Explain how questions with ambiguous answers were handled.\\\hline
\rowcolor{gray!30}\multicolumn{2}{|p{0.97\linewidth}|}{2(iv-a) If human-graded: The evaluation summary briefly describes the sample of graders and how they were recruited.} \\
\hline
\textbf{Minimal Requirements} & \textbf{Full Compliance} \\
\hline
\textbf{2(iv-a)A.} State the domain or other relevant qualifications of graders.\linebreak
\textbf{2(iv-a)B.} Disclose the institutional affiliation of graders.&
\textbf{2(iv-a)C.} State the number of graders.\linebreak
\textbf{2(iv-a)D.} Briefly describe how graders were recruited.\linebreak
\textbf{2(iv-a)E.} WHERE APPLICABLE: Note if graders were provided with training for the grading task.\\\hline
\rowcolor{gray!30}\multicolumn{2}{|p{0.97\linewidth}|}{2(iv-b) If human-graded: The evaluation summary briefly describes the grading materials and process.} \\
\hline
\textbf{Minimal Requirements} & \textbf{Full Compliance} \\
\hline
\textbf{2(iv-b)A.} Describe the content of the grading instructions and rubrics OR provide illustrative examples of grading instructions and rubrics.
\textbf{2(iv-b)B.} State whether graders were blinded to the identity of the test-taker. &
\textbf{2(iv-b)C.} State the typical number of independent graders that graded each item response.\linebreak
\textbf{2(iv-b)D.} Briefly explain the process for adjudicating grader disagreements.\\\hline
\rowcolor{gray!30}\multicolumn{2}{|p{0.97\linewidth}|}{2(iv-c) If human-graded: The evaluation summary describes the level of agreement between graders.} \\
\hline
\textbf{Minimal Requirements} & \textbf{Full Compliance} \\
\hline
\textbf{2(iv-c)A.} Provide some qualitative or quantitative indicator or statement about the level of agreement between graders. &  \textbf{2(iv-c)B.} Provide an appropriate summary statistic for grader agreement (e.g. Cohen’s kappa) OR, if no statistics are suitable, state this and give a brief summary of grader disagreements.\linebreak
\textbf{2(iv-c)C.} WHERE APPLICABLE: Flag grader disagreements with important implications for the capability or risk assessment.\\\hline
\rowcolor{gray!30}\multicolumn{2}{|p{0.97\linewidth}|}{2(v-a) If auto-graded: The evaluation summary identifies the model used as an automated grader and describes any modifications made to it.} \\
\hline
\textbf{Minimal Requirements} & \textbf{Full Compliance} \\
\hline
\textbf{2(v-a)A.} Specify the base model used for grading. &
\textbf{2(v-a)B.} State whether only the base model was used, or if the model was modified for the grading task (e.g. with fine-tuning, task-specific scaffolding, etc).\linebreak
\textbf{2(v-a)D.} WHERE APPLICABLE: Briefly describe any modifications made to the base model for the grading task.\\\hline
\rowcolor{gray!30}\multicolumn{2}{|p{0.97\linewidth}|}{2(v-b) If auto-graded: The evaluation summary briefly describes the automated grading materials and process.} \\
\hline
\textbf{Minimal Requirements} & \textbf{Full Compliance} \\
\hline
\textbf{2(v-b)A.} Provide a brief description of the grading rubrics and grading instructions used OR illustrative examples of grading instructions and rubrics.\linebreak
\textbf{2(v-b)B.} Provide a brief description of how the auto-grader judged performance, e.g. based on similarity with gold standard answers.&
\textbf{2(v-b)C.} Share an example prompt used for the auto-grader (sensitive details can be redacted).\linebreak
\textbf{2(v-b)D.} State whether multiple auto-grader samples were generated per evaluation item response.\linebreak
\textbf{2(v-b)E.} ONLY IF multiple auto-grader samples were generated: State how these scores were aggregated for a final score.\\\hline
\rowcolor{gray!30}\multicolumn{2}{|p{0.97\linewidth}|}{2(v-c) If auto-graded: The evaluation summary states whether the automated grader was validated against human graders or another auto-grader, and if so, reports the level of agreement.} \\
\hline
\textbf{Minimal Requirements} & \textbf{Full Compliance} \\
\hline
\textbf{2(v-c)A.} State whether the auto-grader’s performance was validated against human graders, another auto-grader, or not at all.\linebreak
\textbf{2(v-c)B.} ONLY IF the auto-grader’s performance was validated against human graders: Describe the number of human graders and their qualifications.&
\textbf{2(v-c)C.} Provide a summary statistic for the level of agreement between the auto-grader and the comparison grader; OR, if no comparison was made, provide a brief explanation for why this was not done.\linebreak
\textbf{2(v-c)D.} ONLY IF a comparison between the auto-grader and another grader was made: State whether the comparison was conducted on the full set of evaluation items or a subset.\\\hline
\end{longtable}
}

\subsubsection*{\protect\blank{Model Elicitation}\label{model-elicitation}}

{\renewcommand{\arraystretch}{1.5}
\small
\begin{longtable}{|>{\raggedright}p{0.47\linewidth}|>{\raggedright\arraybackslash}p{0.47\linewidth}|}
\hline
\rowcolor{gray!30}\multicolumn{2}{|p{0.97\linewidth}|}{3(i) The model report specifies which version(s) of the model were tested.} \\
\hline
\textbf{Minimal Requirements} & \textbf{Full Compliance} \\
\hline
\textbf{3(i)A.} Somewhere in the model report, clearly specify which model instance(s) were identical to the final/deployed model (e.g. “launch candidate”); OR make clear that no tested model instance was identical to final version.\linebreak
\textbf{3(i)B.} ONLY IF the evaluation includes any model instances that are not the final/deployed model version: Somewhere in the model report, clearly specify which model instances included in this evaluation had the full deployment set of mitigations/safeguards in place at test time, and which had a reduced/minimal set.&
\textbf{3(i)C.} ONLY IF the evaluation did not include a final/deployed model version: Provide some estimate of the capability difference of at least one of the tested model instances to the final/deployed model. Can be qualitative or quantitative.\linebreak 
\textbf{3(i)D.} Label model instances tested in this evaluation in a way that is clear and consistent with model version descriptions satisfying 3(i)A and 3(i)B.\\\hline
\rowcolor{gray!30}\multicolumn{2}{|p{0.97\linewidth}|}{3(ii) The model report briefly describes all the relevant mitigations active during evaluations, and describes any simulated efforts to circumvent mitigations.} \\
\hline
\textbf{Minimal Requirements} & \textbf{Full Compliance} \\
\hline
\textbf{3(ii)A.} Somewhere in the model report, for either evaluations generally, an applicable subset of evaluations, or this evaluation, briefly list the relevant safeguards and mitigations (e.g. unlearning, safety fine-tuning, content classifiers).\linebreak
\textbf{3(ii)B.} Somewhere in the model report, state whether elicitation conditions included any attempts to bypass active safeguards/mitigations (e.g. jailbreaking attacks); OR, if such attempts were not made, but adversarial use was instead tested using model instances with mitigations/safeguards removed, make this clear by labelling these model instances and displaying their results alongside results for safeguarded model(s).&
\textbf{3(ii)C.} Somewhere in the report, for each specific model instance tested in this evaluation, make clear what set or subset of mitigations/safeguards were in place at test time. (Ex: list uniform set of mitigations applied for ChemBio or automated evals; or, if only testing final/deployed model, state final deployment set.)\linebreak
\textbf{3(ii)D.} Somewhere in the report, briefly describe how rigorous any attempts to bypass active safeguards/mitigations were (e.g. how much time was spent finding jailbreaks); OR, for this evaluation, briefly explain why no bypassing attempts were made (e.g because there were no model refusals).
\textbf{3(ii)E.} IF APPLICABLE: disclose the extent to which model refusals affected evaluation. (Ex: number of items on which refusals occurred.)\\\hline
\rowcolor{gray!30}\multicolumn{2}{|p{0.97\linewidth}|}{3(iii) The model report specifies the actions taken to surface the full range of model capabilities during evaluation.} \\
\hline
\textbf{Minimal Requirements} & \textbf{Full Compliance} \\
\hline
\textbf{3(iii)A.} Somewhere in the model report, briefly describe how models were prompted for evaluations. \linebreak
\textbf{3(iii)B.} Somewhere in the model report, for either evaluations generally, an applicable subsest of evaluations, or this evaluation, state which sampling/generation strategies were used for evaluations. (Ex: “Best-of-5”, “pass@1”, “none”.)\linebreak
\textbf{3(iii)C.} Somewhere in the model report, for either all evaluations, an applicable subset of evaluations, or this evaluation, state whether any tools were provided to the models (e.g. web search, calculators).\linebreak 
\textbf{3(iii)D.} Somewhere in the model report, for either all evaluations, an applicable subset of evaluations, or this evaluation, state whether any scaffolding was used (e.g. agentic scaffolding).\linebreak
\textbf{3(iii)E.} WHERE APPLICABLE: somewhere in the model report, state the use of any fine-tuning of models for evaluations.&
\textbf{3(iii)F.} Somewhere in the model report, briefly describe the prompt design process for evaluations.\linebreak
\textbf{3(iii)G.} IF APPLICABLE: provide examples of prompts used for this evaluation.\linebreak
\textbf{3(iii)H.} Somewhere in the model report, briefly list the tools provided to models for this evaluation; OR state that none were provided.\linebreak
\textbf{3(iii)I.} Somewhere in the model report, briefly describe the scaffolding used for this evaluation; OR state that none was used.\linebreak
\textbf{3(iii)J.} Somewhere in the model report, for either all evaluations, an applicable subset of evaluations, or this evaluation, state what resource ceilings were applied (e.g. maximum inference time/tokens).\linebreak
\textbf{3(iii)K.} Somewhere in the model report, for either all evaluations, an applicable subset of evaluations, or this evaluation, state what sampling parameters were applied (e.g. temperature).\linebreak
\textbf{3(iii)L.} ONLY IF fine-tuning was used (see 3(iii)E): Somewhere in the model report, briefly describe the dataset and/or methods used for fine-tuning.\\\hline
\end{longtable}
}

\subsubsection*{\protect\blank{Model Performance }\label{model-performance}}

{\renewcommand{\arraystretch}{1.5}
\small
\begin{longtable}{|>{\raggedright}p{0.47\linewidth}|>{\raggedright\arraybackslash}p{0.47\linewidth}|}
\hline
\rowcolor{gray!30}\multicolumn{2}{|p{0.97\linewidth}|}{3(i) The model report specifies which version(s) of the model were tested.} \\
\hline
\textbf{Minimal Requirements} & \textbf{Full Compliance} \\
\hline
\textbf{4(i)A.} Present whichever summary statistic(s) for model performance on this evaluation are most appropriate, either in text, or in a figure or graph. &
\textbf{4(i)B.} Clearly present the summary statistic(s) given for 4(i)A either in text, a table, or a graph with clear text labelling (a figure or graph with no numerical labelling of the summary statistic is not sufficient).\linebreak
\textbf{4(i)C.} ONLY IF the summary statistic reported is not mean solve rate or a similar metric: Give a brief justification for the choice of summary statistic(s).\\\hline
\rowcolor{gray!30}\multicolumn{2}{|p{0.97\linewidth}|}{4(ii) The evaluation summary provides confidence intervals (or other uncertainty measures) for performance statistics, and specifies the number of evaluation runs conducted.} \\
\hline
\textbf{Minimal Requirements} & \textbf{Full Compliance} \\
\hline
\textbf{4(ii)A.} Include an appropriate measure of statistical uncertainty for the performance reported for 4(i), e.g. confidence interval, standard error of the mean, either in text, or in a figure or graph. \linebreak
\textbf{4(ii)B.} ONLY IF confidence intervals are given: Include the confidence level (e.g. “95\% CI”).&
\textbf{4(ii)C.} Specify the number of evaluation runs conducted per model that the summary statistics summarize. \linebreak
\textbf{4(ii)D.} Clearly present the uncertainty measure(s) given for 4(ii)A either in text, a table, or a graph with clear text labelling (a figure or graph with no numerical labelling of the uncertainty measure is not sufficient).\\\hline
\rowcolor{gray!30}\multicolumn{2}{|p{0.97\linewidth}|}{4(iii) The evaluation summary states whether ablation experiments or multiple alternative testing conditions were performed, and states whether the model was tested for training contamination.} \\
\hline
\textbf{Minimal Requirements} & \textbf{Full Compliance} \\
\hline
\textbf{4(iii)A.} State whether supplementary evaluation runs were performed with major variations on mainline evaluation conditions (e.g. different elicitation protocols, resource ceilings, or test versions)\linebreak
\textbf{4(iii)B.} ONLY IF supplementary evaluation runs described in 4(iii)A were performed: Report the outcome of each major testing variation (e.g. with summary statistics or a qualitative description).&
\textbf{4(iii)C.} Explicitly confirm whether the model report provides the “highest” score or summary measure on this evaluation that was obtained under any testing condition or variation (where “highest” should be construed as “most concerning”, if numerically higher scores do not indicate more concerning outputs).\linebreak
\textbf{4(iii)D.} State whether the model was tested for contamination of its training data with benchmark content.
4(iii)E. ONLY IF testing for contamination described in 4(iii)D was performed: Briefly summarize the results of this testing.\\\hline
\end{longtable}
}

\subsubsection*{\protect\blank{Baseline Performance }\label{baseline-performance}}

{\renewcommand{\arraystretch}{1.5}
\small
\begin{longtable}{|>{\raggedright}p{0.47\linewidth}|>{\raggedright\arraybackslash}p{0.47\linewidth}|}
\hline
\rowcolor{gray!30}\multicolumn{2}{|p{0.97\linewidth}|}{5(i-a) If human baseline: The evaluation summary states the number of human participants, their qualifications, and how they were recruited.
} \\
\hline
\textbf{Minimal Requirements} & \textbf{Full Compliance} \\
\hline
\textbf{5(i-a)A.} State the total number of human participants for the human baseline test for this evaluation.
\textbf{5(i-a)B.} ONLY IF the report specifies that the human baseline is “expert” level: State the human baseline participants’ specific domain(s) of expertise (e.g. virology) AND their education level or relevant professional experience.
\textbf{5(i-a)C.} ONLY IF 5(i-a)B is not applicable: State the type of human baseline (e.g. “novice”) AND provide some statement about their qualifications, domain knowledge, or other task-relevant characteristics.&
\textbf{5(i-a)D.} Briefly describe how the human baseline sample was recruited (e.g. recruitment channels).
\textbf{5(i-a)E.} WHERE APPLICABLE: Disclose any features of recruitment that were likely to introduce significant sampling bias (e.g. experts all drawn from a single research group).\\\hline
\rowcolor{gray!30}\multicolumn{2}{|p{0.97\linewidth}|}{5(i-b) If human baseline: The evaluation summary provides human performance statistics, and reports any differences between the AI evaluation and human baseline test.} \\
\hline
\textbf{Minimal Requirements} & \textbf{Full Compliance} \\
\hline
\textbf{5(i-b)A.} Present whichever summary statistic(s) for human baseline performance on this evaluation are most appropriate, either in text, or in a figure or graph.& 
\textbf{5(i-b)B.} Include an appropriate measure of statistical uncertainty for the human baseline performance reported for 5(i-b)A, e.g. confidence interval, standard error of the mean, either in text, or in a figure or graph. \linebreak
\textbf{5(i-b)C.} ONLY IF confidence intervals are given: Include the confidence level (e.g. “95\% CI”).\linebreak
\textbf{5(i-b)D.} Clearly present the summary statistic(s) given for 5(i-b)A and the uncertainty measure(s) given for 5(i-b)B either in text, a table, or a graph with clear text labelling (a figure or graph with no numerical labelling of the uncertainty measure is not sufficient).\linebreak
\textbf{5(i-b)E.} ONLY IF the human baseline summary statistic is not either the mean or an identical measure to the model summary statistic in 4(i): Give a brief justification for the choice of human baseline summary measure.\linebreak
\textbf{5(i-b)F.} WHERE APPLICABLE: Report any important differences between the AI evaluation and the human baseline test (e.g. if humans were only graded on questions matching their expertise).\\\hline
\rowcolor{gray!30}\multicolumn{2}{|p{0.97\linewidth}|}{5(i-c) If human baseline: The evaluation summary provides details of the testing conditions in the human baseline experiment.} \\
\hline
\textbf{Minimal Requirements} & \textbf{Full Compliance} \\
\hline
\textbf{5(i-c)A.} Report the amount of time allowed for human baseline participants to complete this evaluation task.\linebreak
\textbf{5(i-c)B.} Describe what resources human participants had access to during the baseline test (e.g. internet access, biological design tools, none).&
\textbf{5(i-c)C.} Briefly describe what incentives participants were given to ensure high motivation for performing well on the test (e.g. hourly base-pay plus performance bonuses).\linebreak
\textbf{5(i-c)D.} State how much time human baseline participants spent on a typical test item, or on the test as a whole, on average.\linebreak 
\textbf{5(i-c)E.} WHERE APPLICABLE: Note any other features of the testing environment that may have significantly impacted performance, or any problems observed at test time (e.g. with motivation or task compliance).\\\hline
\rowcolor{gray!30}\multicolumn{2}{|p{0.97\linewidth}|}{5(ii-a) If no human baseline: The model report explains why a human comparison would not be appropriate or feasible.} \\
\hline
\textbf{Minimal Requirements} & \textbf{Full Compliance} \\
\hline
\textbf{5(ii-a)A.} Briefly explain why including a human baseline for this evaluation would be infeasible (e.g. due to high costs, legal constraints, or safety risks) OR briefly explain why a human baseline for this evaluation would not be informative (e.g. because the test is trivially easy or excessively hard for humans).&
\textbf{5(ii-a)B.} Provide supporting details or evidence for 5(ii-a)A (e.g. authoritative sources consulted, time or cost estimates for human baseline study, supporting research literature).\\\hline
\rowcolor{gray!30}\multicolumn{2}{|p{0.97\linewidth}|}{5(ii-b) If no human baseline: The model report provides an alternative way of interpreting the evaluation in the absence of human comparisons (e.g. an alternative baseline).} \\
\hline
\textbf{Minimal Requirements} & \textbf{Full Compliance} \\
\hline
\textbf{5(ii-b)A.} Provide some other means of interpreting the significance of model performance on this evaluation, such as scores from previously released models, or a summary of expert judgments on appropriate score interpretations for this evaluation.\linebreak
\textbf{5(ii-b)B.} ONLY IF 5(ii-b)A is not met with empirical baselines such as previously released model scores: Briefly describe the methodology for obtaining the expert judgments or other reference point(s) satisfying 5(ii-b)A.&
\textbf{5(ii-b)C.} Justify why the reference point(s) satisfying 5(ii-b)A provide a valid and useful comparison with the main model results, in particular explaining specifically how these reference point(s) could inform an accurate interpretation of a model’s ChemBio capabilities or risk level.\linebreak
\textbf{5(ii-b)D.} Briefly summarize major uncertainties affecting 5(ii-b)A, 5(ii-b)B, or 5(ii-b)C.\\\hline
\end{longtable}
}

\subsubsection*{\protect\blank{Results interpretation }\label{results-interpretation}}

{\renewcommand{\arraystretch}{1.5}
\small
\begin{longtable}{|>{\raggedright}p{0.47\linewidth}|>{\raggedright\arraybackslash}p{0.47\linewidth}|}
\hline
\rowcolor{gray!30}\multicolumn{2}{|p{0.97\linewidth}|}{6(i) The model report states the conclusions the evaluators have drawn about the model’s capabilities and risk level, and connects this with evaluation and other evidence.
} \\
\hline
\textbf{Minimal Requirements} & \textbf{Full Compliance} \\
\hline
\textbf{6(i)A.} Somewhere in the model report, state the overall conclusions drawn about the model’s ChemBio capability level and/or ChemBio risk level.\linebreak
\textbf{6(i)B.} Somewhere in the model report, provide a brief statement on how the conclusion(s) in 6(i)A impacted decision-making (e.g. deployment decisions, level of mitigations, etc.).&
\textbf{6(i)C.} Somewhere in the model report, clearly explain the degree to which specific evaluations contributed to the conclusion(s) in 6(i)A, in one of the following ways: by indicating which evaluations had the most influence on these conclusion(s); OR by indicating which tested capabilities had the most influence (provided these capabilities are clearly tied to specific evaluations); OR by clearly describing a rule or formula used for outputting conclusions from evaluation results.\linebreak
\textbf{6(i)D.} Somewhere in the model report, briefly describe any important influences on the conclusion(s) in 6(i)A \emph{other than} the reported evaluations, e.g. evaluations performed by external parties.\\\hline
\rowcolor{gray!30}\multicolumn{2}{|p{0.97\linewidth}|}{6(ii) The model report states what evidence could have ‘falsified’ the conclusion(s) above, and whether such interpretations were pre-registered in a credible way.} \\
\hline
\textbf{Minimal Requirements} & \textbf{Full Compliance} \\
\hline
\textbf{6(ii)A.} Somewhere in the model report, clearly state what combination of evaluation results or other evidence could have significantly changed the conclusion(s) in 6(i)A—in particular, state what would have resulted in a \emph{higher} risk or capability determination.&
\textbf{6(ii)B.} Somewhere in the model report, state whether the conditions described for 6(ii)A were pre-registered in connection with the higher risk interpretation, either as a public statement or as shared with a credible third party.\\\hline
\rowcolor{gray!30}\multicolumn{2}{|p{0.97\linewidth}|}{6(iii) The model report includes statements about near-term future performance.} \\
\hline
\textbf{Minimal Requirements} & \textbf{Full Compliance} \\
\hline
\textbf{6(iii)A.} Somewhere in the model report, include a statement about how model performance might improve in the near future (3-6 months from release) with further development of elicitation techniques and tools.\linebreak
\textbf{6(iii)B.} ONLY IF the model will be deployed open-source or open-weight: Somewhere in the model report, include a statement about how model performance might improve in the next 12-24 months.\linebreak
\textbf{6(iii)C.} Somewhere in the model report, state any implications of statements for 6(iii)A (and 6(iii)B if applicable) for capability thresholds, risk levels, or mitigations/safeguards.&
\textbf{6(iii)D.} Somewhere in the model report, provide a brief explanation of the statement(s) for 6(iii)A (and 6(iii)B, if applicable).\linebreak
\textbf{6(iii)E.} Somewhere in the model report, provide at least a tentative statement about when an important decision point (e.g. a capability or risk threshold) might be reached by a model in this model family. This can be in terms of calendar time (e.g. “3 months”) or development schedule (e.g. “next major model release”).\\\hline
\rowcolor{gray!30}\multicolumn{2}{|p{0.97\linewidth}|}{6(iv) The model report states how much time the relevant team(s) had to consider evaluation results prior to deployment.} \\
\hline
\textbf{Minimal Requirements} & \textbf{Full Compliance} \\
\hline
\textbf{6(iv)A.} Somewhere in the model report, provide some statement about how long internal safety teams (or whichever groups/individuals are most relevant, such as independent third-party evaluators) had to form and communicate interpretations of evaluation results prior to model deployment.&
\textbf{6(iv)B.} Somewhere in the model report, provide a rough quantified estimate of the time reported in 6(iv)A (e.g. through date ranges, numbers of days, or FT equivalents).\\\hline
\rowcolor{gray!30}\multicolumn{2}{|p{0.97\linewidth}|}{6(v) The model report briefly describes any notable uncertainties or disagreements related to interpreting results or making risk judgments, and how these were handled.} \\
\hline
\textbf{Minimal Requirements} & \textbf{Full Compliance} \\
\hline
\textbf{6(v)A.} Somewhere in the model report, state whether any notable uncertainties or disagreements arose during the ChemBio evaluation and interpretation process. & 
\textbf{6(v)B.} ONLY IF the model report does not explicitly state that there were no uncertainties/disagreements: Somewhere in the model report, briefly summarize notable uncertainties/disagreements (sensitive information can be redacted).\linebreak 
\textbf{6(v)C.} Somewhere in the model report, briefly explain how considerations from 6(v)B were dealt with (e.g. independent review); OR, if there were no uncertainties/disagreements, outline how they would have been addressed, had they occurred.\\\hline
\end{longtable}
}

\subsection*{Terminology:}
\textbf{“Applicable subset of evaluations”} - When criteria refer to information provided for "an applicable subset of evaluations," this includes general statements about evaluation procedures that apply to a broader category or evaluation suite that encompasses the specific evaluation being assessed. For example, if an evaluation is part of the "CBRN evaluations" suite, then general statements about CBRN evaluation methodology would satisfy criteria that allow for "applicable subset" reporting.

\textbf{“State whether”} - The model report must either explicitly state that a given condition was met, explicitly state that it was not met, or provide details of how the condition was met that implicitly confirms it.

\printbibliography
\end{document}